\documentclass[twocolumn]{aastex631}
\usepackage{amsmath}
\usepackage{amssymb}	
\usepackage{gensymb}
\usepackage{gensymb}
\usepackage{graphicx}
\usepackage{array}
\usepackage{xspace}
\usepackage{multirow}
\usepackage{appendix}
\usepackage{fontawesome5}
\usepackage{hyperref}

\newcommand{\lenstro}{\textsc{Lenstronomy}\xspace}
\newcommand{\jax}{\textsc{jax}\xspace}

\newcommand{\starred}{\textsc{starred}\xspace}
\newcommand{\photutils}{\textsc{photutils}\xspace}
\newcommand{\psfr}{\textsc{psfr}\xspace}

\newcommand{\op}[1]{\ensuremath{\boldsymbol{\mathsf{#1}}}\xspace}
\newcommand{\mat}[1]{\ensuremath{{\rm \bf #1}}\xspace}
\newcommand{\bfsym}[1]{\ensuremath{\boldsymbol{#1}}}
\newcommand{\narrowPSF}{\ensuremath{\bfsym{s}(\bfsym{\eta}_{\rm PSF})}\xspace}
\newcommand{\narrowPSFnodep}{\bfsym{s}\xspace}
\newcommand{\largePSF}{\bfsym{t}\xspace}

\newcommand{\deconvnox}{\bfsym{f}\xspace}
\newcommand{\hnox}{\bfsym{h}(\ensuremath{\bfsym{\eta}_{\rm ext}})\xspace}
\newcommand{\largePSFk}{\ensuremath{\bfsym{t}_{ k}}\xspace}
\newcommand{\shiftedgaussian}{\bfsym{r}\xspace}
\newcommand{\shiftedgaussiancentered}{\ensuremath{\bfsym{r}_{ k}}\xspace}

\newcommand{\modelparam}{\ensuremath{{\bfsym{\eta}_{\rm PSF}}}\xspace}
\newcommand{\modelparamdec}{\ensuremath{\bfsym{\eta}_{\rm ext}}\xspace}
\newcommand{\modelparammoffat}{\ensuremath{\bfsym{\eta}_{\rm Moffat}}\xspace}
\newcommand{\modelparamgrid}{\ensuremath{\bfsym{\eta}_{\rm pixel}}\xspace}

\newcommand{\data}{\bfsym{d}}

\newcommand{\subop}{\op{D}}
\newcommand{\convop}{\ensuremath{\op{B}_{ k}}\xspace}
\newcommand{\sconvop}{\op{S}}
\newcommand{\noise}{\bfsym{n}}
\newcommand{\covmatrix}{\mat{C}}

\newcommand{\waveletop}{\ensuremath{\bfsym{\Phi}}\xspace}

\newcommand{\weights}{\ensuremath{\mat{W}}\xspace}

\newcommand{\prior}[1]{\ensuremath{\mathcal{P}\left(#1\right)}\xspace}

\newcommand{\loss}{\ensuremath{L}\xspace}

\newcommand{\likelihood}[1]{\ensuremath{\mathcal{L}\left(#1\right)}\xspace}

\newcommand{\priorsimple}{\ensuremath{\mathcal{P}}\xspace}

\newcommand{\normone}[1]{\ensuremath{\left\lVert \,#1\, \right\rVert_1}}
\newcommand{\normtwo}[1]{\ensuremath{\left\lVert \,#1\, \right\rVert_2}}

\newcommand{\Pastro}{\ensuremath{\mathrm{P}_{\rm astro}}\xspace}

\newcommand{\MAEPSF}{\ensuremath{\mathrm{MBE}_{\rm PSF}}\xspace}

\DeclareMathOperator*{\argmin}{arg\,min}

\begin{document}

\title{Image deconvolution and PSF reconstruction with STARRED: a wavelet-based two-channel method optimized for light-curve extraction}

\author[0000-0001-7051-497X]{Martin Millon}
\affiliation{Kavli Institute for Particle Astrophysics and Cosmology and Department of Physics, Stanford University,
Stanford, CA 94305, USA \label{KIPAC}\\}
\author[0000-0002-6584-2749]{Kevin Michalewicz}
\affiliation{Department of Mathematics, Imperial College London, London SW7 2AZ, United Kingdom \label{ICL}\\}
\author[0000-0003-3358-4834]{Fr\'ed\'eric Dux}
\affiliation{European Southern Observatory,
Alonso de Córdova 3107,
Vitacura, Santiago, Chile
\label{ESO}}
\affiliation{Institute of Physics, Laboratory of Astrophysics, \'Ecole Polytechnique F\'ed\'erale de Lausanne (EPFL), Observatoire de Sauverny, 1290 Versoix, Switzerland 
\label{epfl} }
\author[0000-0003-0758-6510]{Fr\'ed\'eric Courbin}
\affiliation{Institute of Physics, Laboratory of Astrophysics, \'Ecole Polytechnique F\'ed\'erale de Lausanne (EPFL), Observatoire de Sauverny, 1290 Versoix, Switzerland 
\label{epfl} }
\author[0000-0002-0113-5770]{Philip J. Marshall}
\affiliation{Kavli Institute for Particle Astrophysics and Cosmology and Department of Physics, Stanford University,
Stanford, CA 94305, USA \label{KIPAC}\\}
\affiliation{SLAC National Accelerator Laboratory, Menlo Park, CA 94025, USA \label{SLAC}\\}

\begin{abstract}
We present \starred, a Point Spread Function (PSF) reconstruction, two-channel deconvolution, and light curve extraction method designed for high-precision photometric measurements in imaging time series. An improved resolution of the data is targeted rather than an infinite one, thereby minimizing deconvolution artifacts. In addition, \starred performs a joint deconvolution of all available data, accounting for epoch-to-epoch variations of the PSF and decomposing the resulting deconvolved image into a point source and an extended source channel. The output is a high signal-to-noise, high resolution frame combining all data, and the photometry of all point sources in the field of view as a function of time. Of note, \starred also provides exquisite PSF models for each data frame. We showcase three applications of \starred in the context of the imminent LSST survey and of \textit{JWST} imaging: i) the extraction of supernovae light curves and the scene representation of their host galaxy, ii) the extraction of lensed quasar light curves for time-delay cosmography, and iii) the measurement of the spectral energy distribution of globular clusters in the ``Sparkler", a galaxy at redshift z=1.378 strongly lensed by the galaxy cluster SMACS J0723.3-7327. \starred is implemented in \jax, leveraging automatic differentiation and GPU acceleration. This enables rapid processing of large time-domain datasets, positioning the method as a powerful tool for extracting light curves from the multitude of lensed or unlensed variable and transient objects in the Rubin-LSST data, even when blended with intervening objects. 

\end{abstract}


\section{Introduction} 
\label{sec:intro}

The resolution of astronomical images is fundamentally limited by the atmospheric turbulence in the Earth's atmosphere or by optical diffraction of the telescope in the case of space observations. Both of these effects induce blurring of the images, which can be corrected by knowing the Point Spread Function (PSF), a linear mapping between the flux intensity in the sky and that actually recorded at the focal plane of the instrument. However, the PSF evolves temporally with changing atmospheric conditions.

Temporal variations of the PSF reduce the photometric accuracy that can be achieved for blended sources, since light from adjacent sources is mixed differently in changing seeing conditions. This introduces systematic errors in the light curves that, unlike photon noise, cannot be reduced with longer exposures (systematics in fact actually get worse with increasing signal-to-noise). This typically affects lensed quasars, as the light of the lensed images is often blended with the foreground lensing galaxy. 

The epoch-to-epoch variations of the PSF can be corrected by modeling the PSF of each individual image and then removing its effect from the data, a problem known as PSF deconvolution. This inverse problem is mathematically ill-posed due to pixelization and presence of noise, resulting in the absence of a unique and stable solution. To address this, one can introduce regularization to the solution, essentially acting as a Bayesian prior that constrains the parameter space for potential solutions. Regularization selects the best solution among the plethora of possible ones, given some prior information embedded in the regularization term.

Historically, deconvolution techniques such as Tikhonov's~\citep{Tikhonov1977}, characterized by a quadratic regularization term and a closed-form solution, and its related Wiener's deconvolution~\citep{Sekko1999}, have been pivotal. Others have postulated that the optimal solution must be the one that has maximum entropy~\citep{Skilling1984, Narayan1986}.
CLEAN~\citep{Hogbom1974, Cornwell2008}, a well-known early method to image deconvolution for sky brightness estimation, was mostly employed in radio astronomy. With the arrival of the Hubble Space Telescope, the Richardson-Lucy algorithm~\citep{Richardson1972, Lucy1974} became popular. It implements an iterative process based on a Bayesian approach to estimate the most likely model from a noisy image, accounting for the influence of the PSF. This algorithm was later extended to a two-channel decomposition in~\cite{Pirzkal2000},~\cite{Becker2003} and~\cite{Velusamy2008}, specifically designed for separating point sources and extended sources. Subsequent advancements involved introducing prior assumptions in the form of a hierarchical parameter model~\citep{Selig2015}. 

The advent of Deep Neural Networks (DNNs) has then improved performance and inference times by leveraging existing databases and learning features in a supervised fashion~\citep{Sureau2020, Akhaury2022, Nammour2022, Adam2023}. To date, all of these methods have attempted to correct for the full PSF, resulting in the Nyquist-Shannon sampling theorem not being satisfied~\citep{Magain1998}: a deconvolved point source, which is essentially a Dirac function, can never be accurately represented by any sampling or pixel size. Deconvolving by the full PSF is therefore prone to producing unwanted image artifacts (``ringing'') since the deconvolved image is to be represented on an array of pixels, regardless of its size. In general these artifacts can be suppressed by the inclusion of prior information, either in the form of a regularization term in a likelihood-based approach, or via the training sets for the Deep Learning-based methods. Typically a DNN will require a very large training set of images that have better resolution than the data, which is challenging to generate. For these reasons, both Deep Learning and standard approaches should be considered complementary. 

Addressing this challenge of partial deconvolution, Firedec~\citep{Cantale2016} proposes a two-channel method to improve the resolution of the data rather than completely eliminating the effect of the PSF. Specifically, it decomposes the images into a parametric point source channel and an extended source channel, represented on a pixel grid and enforcing wavelet regularization. However, the Haar wavelets considered in Firedec are not shift-invariant and are not well-adapted to the typical topology of astronomical objects.

Wavelet regularization introduces a minimally-informative prior to penalize the presence of non-zero elements in the wavelet domain. An isotropic wavelet basis in which astronomical objects are well represented, i.e. \textit{sparsely} represented, is the starlets~\citep{Starck2015}, meaning that they can be expressed as a linear combination of a few atoms from the starlet dictionary. A sparse representation of the signal is a desired property as it facilitates image denoising. Since the noise is not sparse in the starlet domain, it can be reduced by reconstructing the image only from the most significant coefficients, in a technique called \textit{image thresholding}. The different scales resulting from the starlet transform can be penalized differently, with the noise being mostly represented in the finer wavelet scale.

In this paper, we present the mathematical framework as well as several tests and science applications of
\starred, a method we previously introduced in~\cite{Michalewicz2023}.
\starred 
performs both the PSF reconstruction and the image deconvolution steps, based on the principles of MCS~\citep{Magain1998} and Firedec~\citep{Cantale2016}, but with various key enhancements. These include the use of a family of isotropic wavelets (starlets) 
specifically designed to represent astronomical objects, automatic differentiation, accelerated execution, scalability on Graphics Processing Units (GPUs), and the capability to handle multi-epoch deconvolution. 

The method is very general and has a wide range of applications. 
These include but are not limited to light curve extraction of variable objects such as Active Galactic Nuclei (AGN) or supernovae, point source photometry in crowded fields, or the quasar/host decomposition problem~\citep[see e.g.,][for recent comparison between existing methods]{Leist2023}. 
Here, we select three science cases to demonstrate the capabilities of \starred: i) light curve extraction and scene modeling of Type Ia Supernovae, ii) time-delay measurement in lensed quasars and iii) Spectral Energy Distribution (SED) measurement of high-redshift globular clusters. 
We apply this method to simulated Rubin-LSST time series and to real \textit{JWST} observations, assessing the performance of the algorithm in terms of photometric and astrometric accuracy, and computational speed. 

The paper is organized as follows: in Section~\ref{sec:formalism} we present the mathematical formalism of the method. In Sections~\ref{sec:PSF_performance} and~\ref{sec:deconvolution_performance}, we assess the performance of the PSF reconstruction and image deconvolution methods respectively\footnote{Examples and tests presented in this paper can be reproduced from this repository: \url{https://gitlab.com/cosmograil/starred-examples}}. We discuss our results in Section~\ref{sec:discussion} and conclude in Section~\ref{sec:conclusion}, also mentioning the limitations and future improvements of \starred. 

\section{Formalism} 
\label{sec:formalism}

\begin{figure}[h!]
    \centering
    \includegraphics[width=0.47\textwidth]{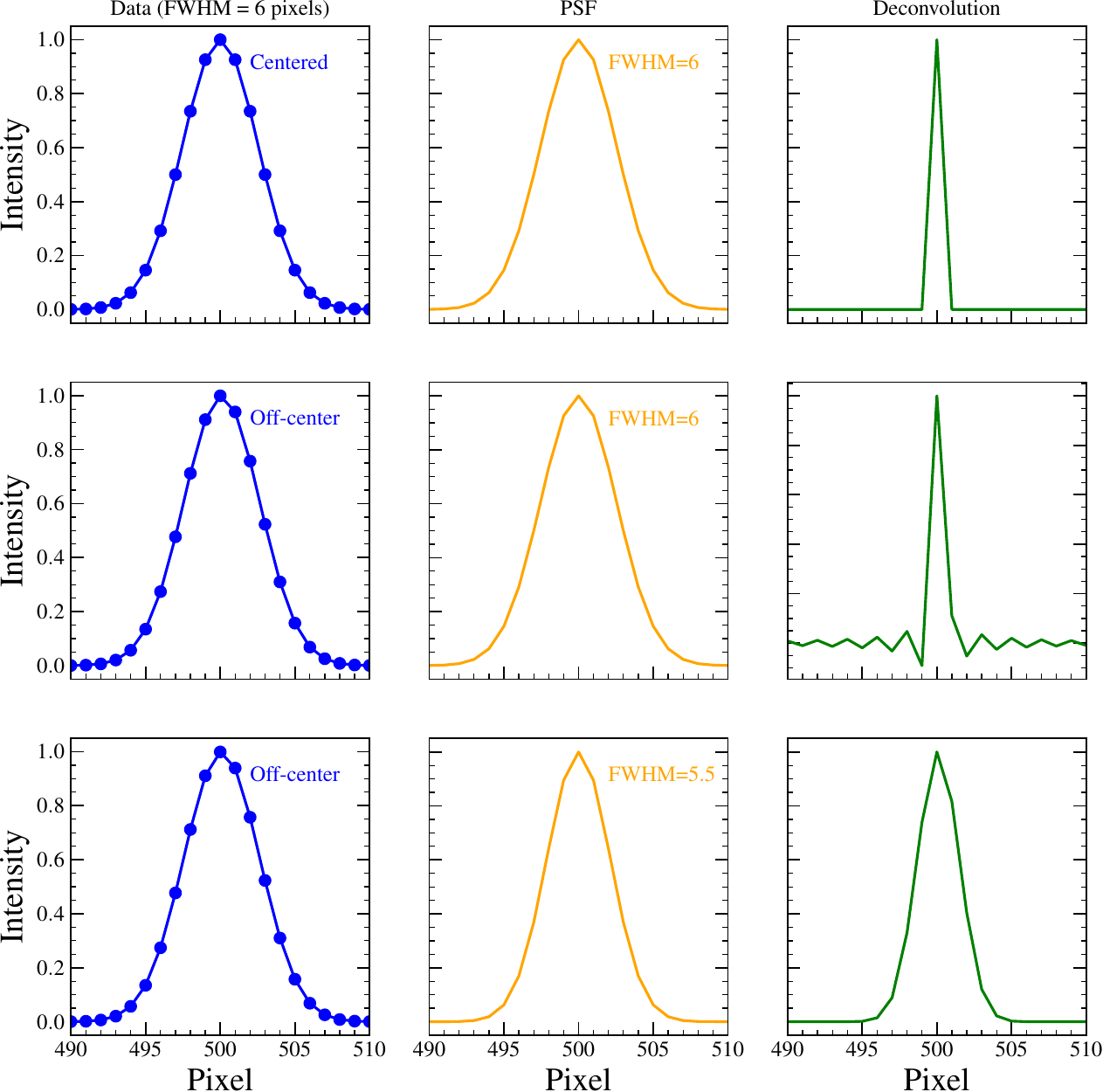}
    \caption{Illustration of Gibbs oscillations due to aliasing of high frequencies in the frequency domain allowed by the pixel size. {\it Top:} from left to right, a noise-free point source is perfectly centered on a pixel and is deconvolved with a PSF that has the same FWHM. Here we use direct Fourier division to do the deconvolution as there is no noise involved. The deconvolved point source is a Delta function also perfectly centered on a pixel. All frequencies, even infinite, are well represented regardless of the pixel size, i.e., no artifacts appear. {\it Middle:} exact same experiment but now the signal is off-centered by 0.1 pixel with respect to the center of a pixel. High frequencies are not well represented and Gibbs oscillations do appear. Note also that: i) the mean flux level is now wrong, ii) the impact of Gibbs oscillations depends on where a given star in the data falls with respect to the pixels, with the worst effect corresponding to stars appearing exactly between two pixels. {\it Bottom:} same as in the middle row but now the PSF used to deconvolve is narrower than the data, hence the deconvolved image does not contain infinitely high frequencies that are impossible to sample. The deconvolved point source has a finite resolution and does not present Gibbs oscillations.}
    \label{fig:aliasing}
\end{figure}

The maximum resolution that a deconvolution algorithm can achieve is fundamentally limited by the Nyquist-Shannon sampling theorem, which states that for an evenly sampled signal, and in the absence of any additional information, no frequencies higher than twice the sampling frequency can be recovered. 
Historically, deconvolution algorithms ignored this theorem, by deconvolving with the observed PSF, hence attempting to recover an infinitely high resolution. 
The result of this deliberate choice of going beyond the Nyquist-Shannon limit is illustrated in one dimension in Figure~\ref{fig:aliasing}. 
The deconvolved image is free of artifacts only if the signal is perfectly centered on the pixel grid, which is the only configuration allowing to sample a Dirac function. 
As soon as the signal is (even slightly) off-centered with respect to a pixel, aliasing immediately appears in the deconvolved output. 
However, if a narrower PSF is chosen for deconvolution, as illustrated on the last row of Figure~\ref{fig:aliasing}, the resulting output satisfies the Nyquist-Shannon limit and the signal is reconstructed without artifacts. 
This remains valid in two dimensions when considering image deconvolution. 

\subsection{PSF reconstruction}
\label{subsec:PSFrec}
To circumvent the issue mentioned above, \starred follows MCS and Firedec in not attempting to completely remove the PSF effect when reconstructing images on pixel grids. It instead proposes to decide in advance the resolution and even the shape of the PSF in the deconvolved image, ensuring that it is well sampled given the pixel size adopted to represent the deconvolved image. Of course, this pixel size can well be smaller than in the original data so that the gain in resolution can be very effective even if the sampling of the original data is not good (e.g., \textit{HST} or \textit{JWST} images). The choice made in \starred is that the PSF in the deconvolved image is an isotropic Gaussian with two pixels Full Width Half Maximum (FWHM). This choice, which adheres to the Nyquist-Shannon sampling theorem, involves reconstructing a so-called \textit{narrow} PSF \narrowPSFnodep, which relates to the original PSF \largePSF through: 

\begin{equation}
    \largePSF = \left( \narrowPSFnodep * \shiftedgaussian \right)^\downarrow,
\end{equation}
where \shiftedgaussian is a Gaussian with 2 pixels FWHM and $\downarrow$ performs the subsampling operation~\citep{Magain1998}. We decompose \narrowPSFnodep as a sum of an analytical Moffat function $\bfsym{m} (\modelparammoffat)$ and a grid of pixels $\bfsym{b} (\modelparamgrid)$ such that:
\begin{equation}
    \narrowPSF = \bfsym{m} (\modelparammoffat) +  \bfsym{b} (\modelparamgrid),
\end{equation}
where $\modelparam = (\modelparammoffat, \modelparamgrid)$ are the free parameters of the model. 
This narrow PSF is the one used in all the rest of the work, in replacement of the observed PSF and can be constructed from stars in the field of view. 
In Figure~\ref{fig:aliasing}, $\bfsym{t}$ corresponds to the left column, $\bfsym{s}$ corresponds to the middle column, and $\bfsym{r}$ is the right column. In other words, $\bfsym{s}$ is the kernel that transforms $\bfsym{t}$ into $\bfsym{r}$, where the latter is an isotropic Gaussian profile of any chosen FWHM. From a star cutout \largePSFk containing noise \noise, one can solve the following inverse problem to obtain an estimate of \narrowPSFnodep: 

\begin{equation}
    \largePSFk =  \left(\narrowPSFnodep * a_k \shiftedgaussiancentered \right)^\downarrow + \noise, 
\end{equation}
where \shiftedgaussiancentered is a shifted Gaussian such that $\bfsym{r}_k(\bfsym{x}) = \bfsym{r}(\bfsym{x} - \bfsym{c}_k)$ with $\bfsym{c}_k$ the center of the Gaussian and $a_k$ its amplitude. By defining \convop as the (shifted) Gaussian convolution operator for the $k^{\mathrm{th}}$ star and \subop as the downsampling operator, we write the data fidelity term as the generalized $\chi^2$ of the model for a set of $N$ stars:

\begin{multline}
    \label{eq:likelihoodPSF}
    \likelihood{\largePSF\,|\,\modelparam, a_k, \bfsym{c}_k} =
    \frac12\, \sum_{k=1}^{N} \Big[\,\largePSFk - \subop\,\convop (a_k, \bfsym{c}_k)\,\narrowPSF \Big]^\top \\ 
    \covmatrix^{-1} \Big[\,\largePSFk - \subop\,\convop (a_k, \bfsym{c}_k)\,\narrowPSF \Big],
\end{multline}
where $\covmatrix^{-1}$ is the inverse of the noise covariance matrix. Here, we assume spatially uncorrelated noise, so $\covmatrix$ is a diagonal matrix. Of course, nothing prevents from applying \starred to data with correlated noise, by using the full noise covariance matrix. This problem is unstable in the presence of noise in the data, which requires adding some prior information on the model parameters \modelparamgrid to constraint the solution. The loss function \loss thus reads 
\begin{align}
    \label{eq:loss_function}
    \loss(\modelparam, a_k, \bfsym{c}_k) = \likelihood{\largePSF |\modelparam, a_k, \bfsym{c}_k} + \prior{\modelparamgrid},
\end{align}
the log-likelihood first term capturing the data fidelity, and the second a prior or regularization term. The latter is acting in our case only on a subset of the parameters \modelparamgrid and is defined as:
\begin{equation}
    \label{eq:reg_term}
    \priorsimple (\modelparamgrid) = \lambda \,\normone{ \weights \odot \waveletop^\top\, \bfsym{b} (\modelparamgrid)}, 
\end{equation}
where $\normone{\cdot}$ corresponds to the $\ell_1$-norm, $\lambda$ is the regularization strength, $\waveletop$ is the starlet transform operator, and $\odot$ is the Hadamard product, i.e., the element-wise product. This prior is equivalent to applying a soft-thresholding of the starlet coefficients with a level based on the hyperparameter $\lambda$, expressed in noise standard deviation units \citep{Starck2016}. The matrix $\weights$ weights each starlet coefficient by its corresponding noise (see Section~\ref{subsec:Noise_prop} and~\cite{Joseph2019} for more details). $\lambda$ is crucial hyperparameter to tune the strength of the regularization. It is often set between 1 and 3 to ensure a high-significance of the reconstructed signal. Setting $\lambda = 3$ means that only the coefficient exceeding the noise level by $3\sigma$ will be reconstructed.

Overall, the minimization problem under the above sparsity constraint reads

\begin{multline}
    \label{eq:argmin_PSF_full}
    \argmin_{\modelparam, a_k, \bfsym{c}_k} \frac12\, \sum_{k=1}^{N} \Big[\,\largePSFk - \subop\,\convop (a_k, \bfsym{c}_k)\,\narrowPSF \Big]^\top \covmatrix^{-1}\ \\
    \Big[\,\largePSFk - \subop\,\convop (a_k, \bfsym{c}_k)\,\narrowPSF \Big] \\
    + \lambda\,\normone{ \weights \odot \waveletop^\top\, \bfsym{b}(\modelparamgrid)}.
\end{multline}

\subsection{Two-channel multi-epoch deconvolution}

Given the \textit{narrow} PSF at the $i^{\rm th}$ epoch, $\narrowPSFnodep _i$, the single-epoch deconvolution problem can be formulated as follows: 
\begin{equation}
    \data _i =  (\narrowPSFnodep _i * \deconvnox _i) ^\downarrow + \noise _i \ , 
\end{equation}
where $\data _i$ is the original image, $\deconvnox _i$ is the deconvolved higher-resolution image, and $\noise _i$ is the noise. In \starred, $\deconvnox _i$ is modeled as the sum of the contributions of two distinct \textit{channels}, a point source channel and an extended source channel:

\begin{equation}
    \deconvnox _i (\modelparamdec, \bfsym{c}_k, a_{ki},\Gamma_i) = \Gamma_i + \hnox + \sum_{k=1}^{M} a_{ki} \bfsym{r}_k \ .
\end{equation}

The point source channel consists of a sum of $M$ Gaussian functions $\bfsym{r}_k$, whose amplitudes $a_{ki}$ can vary between epochs, unlike their centers $\bfsym{c}_k$, which are fixed for all epochs. The extended source channel consists of a grid of pixels \hnox common to all epochs. Each pixel contained in \modelparamdec is a free parameter. $\Gamma_i$ accounts for imperfect sky subtraction by applying a spatially constant background correction. 

Similarly to Equation~\ref{eq:likelihoodPSF}, the data fidelity term reads 
\begin{multline}
    \label{eq:likelihood_deconv}
     \likelihood{\data|\modelparamdec, a_{ki}, \bfsym{c}_k, \Gamma_i} = \\
     \frac12 \, \sum_{i=1}^{n_{\rm epoch}} \Big[\,\data _i - \subop\,\sconvop _i \deconvnox _i \Big]^\top \covmatrix^{-1}\Big[\,\data _i - \subop\,\sconvop _i \deconvnox _i  \Big] \ ,
\end{multline}

where we have defined $\sconvop _i$ as the convolution by $\narrowPSFnodep _i$, the narrow PSF kernel corresponding to the $i^{\rm th}$ epoch. Similarly to the PSF reconstruction case, we apply a sparsity prior to regularize the extended channel contribution and the point source channel to avoid flux leakage from the extended channel to the point source channel. The deconvolution minimization problem can then be written as: 

\begin{multline}
    \label{eq:argmin_deconv_full}
    \argmin_{\modelparamdec, a_{ki}, \bfsym{c}_k, \Gamma_i} \frac12\, \sum_{i=1}^{n_{\rm epoch}} \Big[\,\data _i - \subop\,\sconvop _i \deconvnox _i \Big]^\top \covmatrix^{-1}\ \Big[\,\data _i - \subop\,\sconvop _i \deconvnox _i  \Big] \\  
    + \lambda\,\normone{ \weights \odot \waveletop^\top\, \hnox} \\
    + \lambda_{\rm pts} \normone{ \weights \odot \waveletop^\top\, \sum_{k=1}^{M} \langle a_{k} \rangle \bfsym{r}_k}, 
\end{multline}

where  $\langle a_{k} \rangle$ is the mean amplitude over all epochs of the $k^{\rm th}$ point source. If the data are sufficiently informative to lift the degeneracy between the point source and the background, $\lambda_{\rm pts}$ can be set to 0 to avoid the unnecessary computation of the second regularization term. 
This scenario occurs, for instance, when the point sources fade and are no longer detectable, as demonstrated in our example in Section~\ref{subsec:SNIa}. However, in situations where there is significant blending between the point source and the extended source and this degeneracy remains for all epochs, it becomes necessary to include this correction term. $\lambda_{\rm pts}$ can then be estimated iteratively as indicated in Appendix~\ref{ap:C}.


\subsection{Noise propagation}
\label{subsec:Noise_prop}
The weight matrix $\weights$ in Equation~\ref{eq:argmin_deconv_full} plays a central role in controlling the regularization strength as it re-weights the starlet coefficients according to their corresponding noise levels in the direct space. In other words, the coefficients of $\weights$ are derived from the noise contained in the original images once propagated to the starlet coefficients of $\bfsym{h}$ (or $\bfsym{b}$ for the PSF formulation). This implies applying the transposed convolution operators to several noise realizations, before empirically evaluating the variance of the starlet coefficients \citep{Joseph2019, Galan2022}. 

More specifically, a noise realization $\delta\bfsym{d}$ according to an upsampled version of the noise covariance matrix $\covmatrix$, denoted as $\covmatrix^\uparrow$, is propagated to the grid of pixels as $\sconvop_1^\top(\covmatrix^\uparrow)^{-1}\delta\bfsym{d}$, and as $\waveletop^\top\sconvop_1^\top(\covmatrix^\uparrow)^{-1}\delta\bfsym{d}$ in the starlet space. In practice, we use a Monte Carlo approach to generate several realizations of $\delta\bfsym{d}$ and calculate the standard deviations of the starlet coefficients within each starlet scale. These correspond to the coefficients of the $\weights$ matrix. Note that, for multi-epoch deconvolution (or multi-star PSF reconstruction), the noise realization $\delta\bfsym{d}$ can be drawn from the \textit{effective} noise map, which corresponds to the noise map of the co-added data. The regularization strength hyperparameter $\lambda$ is then expressed in unit of noise standard deviation for the combined dataset. For an illustration of this procedure, we refer the reader to Appendix \ref{ap:B}.

\begin{figure}[h!]
    \centering
    \includegraphics[width= 0.5\textwidth]{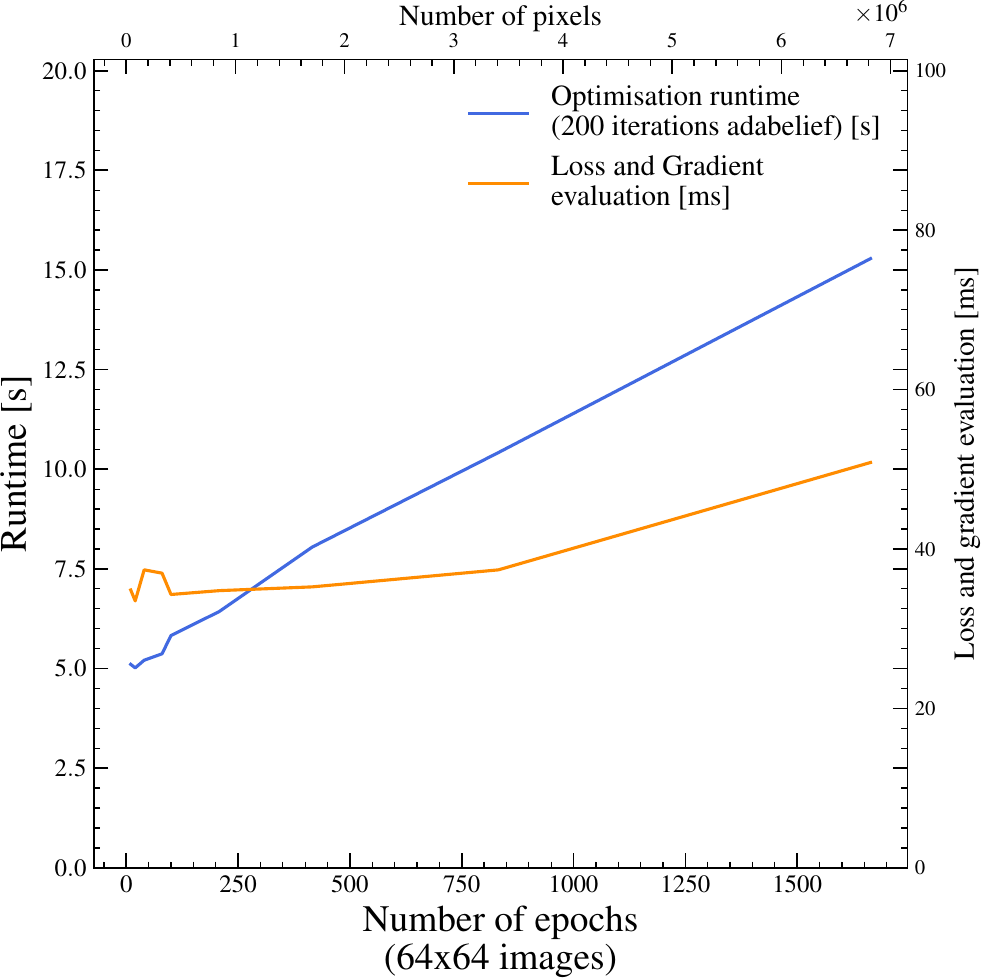}
    \includegraphics[width= 0.5\textwidth]{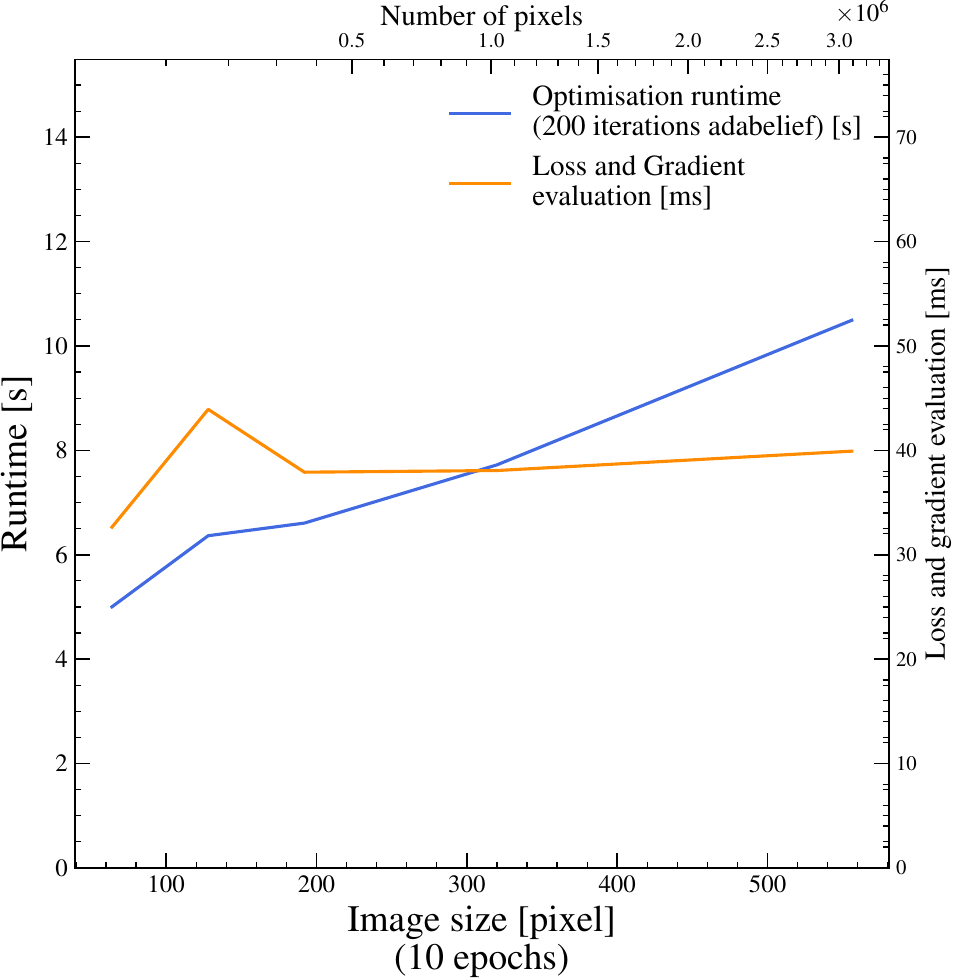}
    \caption{\textit{Upper panel:} \starred's deconvolution runtime on a NVIDIA~RTX~3060 GPU for the deconvolution of 32 $\times$ 32 input images (64 $\times$ 64 deconvolved images) as a function of the number of epochs. The full optimization runtime for 200 iterations of {\tt adabelief} is shown in blue (in seconds), whereas the time needed for a single evaluation of the loss and its gradient is shown in orange (in milliseconds). \textit{Lower panel:} \starred deconvolution runtime for 10 epochs as a function of the output image size. These plots can be reproduced from this notebook \href{https://gitlab.com/cosmograil/starred/-/blob/main/profiling/benchmark_deconvolution/deconv.ipynb?ref_type=heads}{\faGitlab}.
} 
    \label{fig:benchmark_gpu}
\end{figure}

\subsection{JAX implementation}
\label{subsec:jax}

\starred proposes a \jax implementation for the PSF reconstruction (Equation~\ref{eq:argmin_PSF_full}) and deconvolution (Equation~\ref{eq:argmin_deconv_full}) optimization problems. We use \jax automatic differentiation to compute the gradient of the loss function and accelerate the convergence by using a gradient-informed optimizer, such as \texttt{adabelief} or \texttt{l-bfgs-b} from the \textsc{optax} and \textsc{jaxopt} library~\citep{optax, jaxopt}. \starred supports hardware acceleration on GPU. The deconvolution runtime on a single consumer-grade GPU (NVIDIA~RTX~3060) as a function of the input vector size is shown in Figure~\ref{fig:benchmark_gpu}. 

In the context of the large volume of data produced by LSST, \starred will be capable of deconvolving and extracting the light curve for the nearly 1 000 epochs collected during the 10-year survey in approximately 10 seconds, utilizing just a single GPU (assuming 32 $\times$ 32 pixels input cutouts, deconvolved to 64 $\times$ 64 images). 
These 10 seconds include a compilation time of about 2-3 seconds, which is necessary only for the first run and can be avoided in any following deconvolution with the same input size. In this example, the runtime is about 30 times slower on a CPU. 

Unlike deconvolution, PSF reconstruction cannot be performed for all the data at once and needs to be performed sequentially, as the PSF is expected to vary from epoch to epoch. However, this is easily parallelizable. Reconstructing the PSF for 6 $\times$ 32$\times$ 32 star cutouts takes approximately 2.5 seconds on a GPU. If the method is deployed on a large scale, the processing speed may not be sufficient for handling millions of cutouts across thousands of epochs in the LSST survey. The main bottleneck in this process is not so much the optimization runtime, which fully leverages GPU capabilities, but rather the input/output (I/O) and pre-processing times, which are CPU-dependent. However, the I/O runtime and the CPU/GPU memory transfer could be largely optimized on a high-performance computing cluster. Moreover, if computational speed remains a concern, a feasible alternative would be to use the pre-calculated PSF models that are provided as part of the LSST image data products, reserving the PSF reconstruction method solely for the highest priority light curve extractions. However, this may come with a loss in photometric accuracy, if the PIFF PSF modeling approach that is planned for the LSST image processing pipeline follows that of the Dark Energy Survey~\citep{PIFF}. Further investigation is needed but this might just be a necessary trade-off for certain science cases. 

In summary, {\starred} is already scalable for extracting the photometry of millions of transients found in the LSST survey if an accurate PSF model for each epoch is provided. Using the code to reconstruct the PSF for all these cutouts is likely achievable, but may still require some optimization. 

\section{PSF reconstruction performance}
\label{sec:PSF_performance}
\subsection{Metric definitions}
\label{subsec:metrics_def}

\begin{figure*}[h!]
    \centering
    \includegraphics[width=0.9\textwidth]{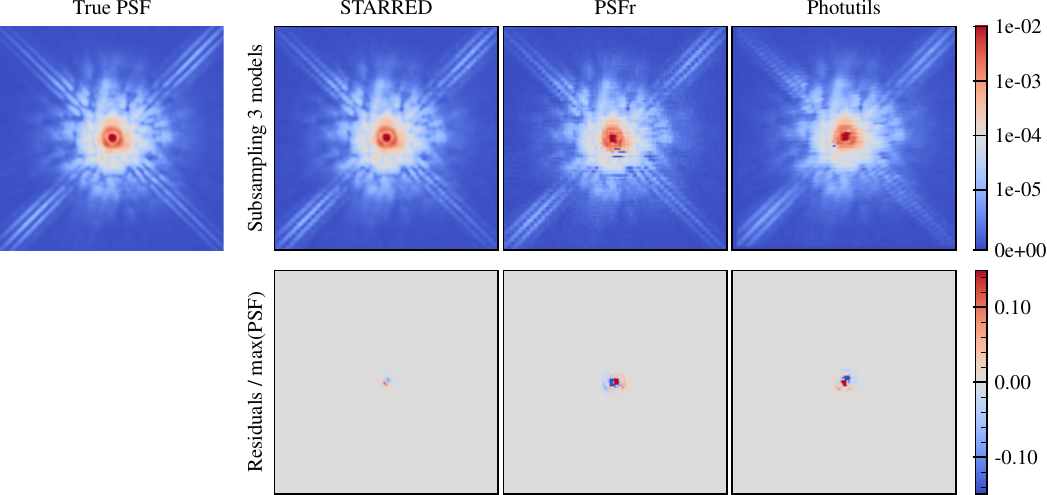}
    \caption{PSF reconstruction of the TinyTim PSF from 6 simulated stars with {\starred}, \psfr and \photutils (top row). The PSF is reconstructed  with a pixel size three times smaller than that of the original data. The bottom row shows the difference between the reconstructed PSFs and the ground truth. This figure can be reproduced from this notebook \href{https://gitlab.com/cosmograil/starred/-/blob/main/notebooks/more_examples/6_HST-PSF\%20reconstruction.ipynb?ref_type=heads}{\faGitlab}. 
    }
    \label{fig:HST_PSF}
\end{figure*}

To evaluate the performance of the PSF reconstruction algorithm, we define 4 metrics, quantifying the astrometric precision, the photometric accuracy and precision and the fidelity of the PSF reconstruction.

We define the astrometric precision as: 

\begin{equation}
    \mathrm{P}_{\rm astro}= \frac{1}{N} \sum_{k=1}^{N} \normtwo{\bfsym{c}_k - \bfsym{\hat{c}}_k - \bfsym{b}}, 
\end{equation}
where $\bfsym{\hat{c}}_k$ is the estimated astrometric position of the k$^{\rm th}$ star and $\bfsym{c}_k$ is its input position in the simulation. $\bfsym{b}$ is the astrometric bias defined as:  
\begin{equation}
   \bfsym{b}= \frac{1}{N}\sum_{k=1}^{N} (\bfsym{c}_k -  \bfsym{\hat{c}}_k).
\end{equation}
We focus mostly on the astrometric precision since a subpixel bias can be introduced during the forward simulation process while interpolating the data on a pixel grid.

The relative photometric accuracy is defined as 

\begin{equation}
    \mathrm{A}_{\rm photo} = \frac{1}{N} \sum_{k=1}^{N} \frac{f_k - \hat{f}_k }{\hat{f}_k}, 
\end{equation}

while the relative photometric precision reads

\begin{equation}
    \mathrm{P}_{\rm photo} = \frac{1}{N} \sum_{k=1}^{N} \bigg | 
 \frac{f_k - \hat{f}_k}{\hat{f}_k} - \mathrm{A}_{\rm photo}\bigg |. 
\end{equation}

We also define a metric to quantify the fidelity of the PSF reconstruction, i.e., the Mean Absolute Error (MAE) of the PSF:

\begin{equation}
    \mathrm{MAE}_{\rm PSF} =  \normone{\bfsym{\mathrm{PSF}} - \widehat{\bfsym{\mathrm{PSF}} }}.
\end{equation}

\subsection{Results}
In this first test, we simulated 6 high-SNR star cutouts as observed by the Hubble Space Telescope (HST) in the F814W filter. These mock observations were generated from the TinyTim PSF~\citep{Krist2011}, subsampled by a factor of 10. We chose a random sub-pixel shift to mimic different star positions relative to the pixel grid and we then interpolated the 10 times sub-sampled TinyTim PSF before downsampling the simulated images by a factor of 5, to obtain a final pixel size of 0.04\arcsec. Finally, we added a background and Poisson noise to the image and adjusted the amplitude of the stars to obtain a SNR of 800 000 for each star. 

\begin{figure*}[h!]
    \centering
    \includegraphics[width=\textwidth]{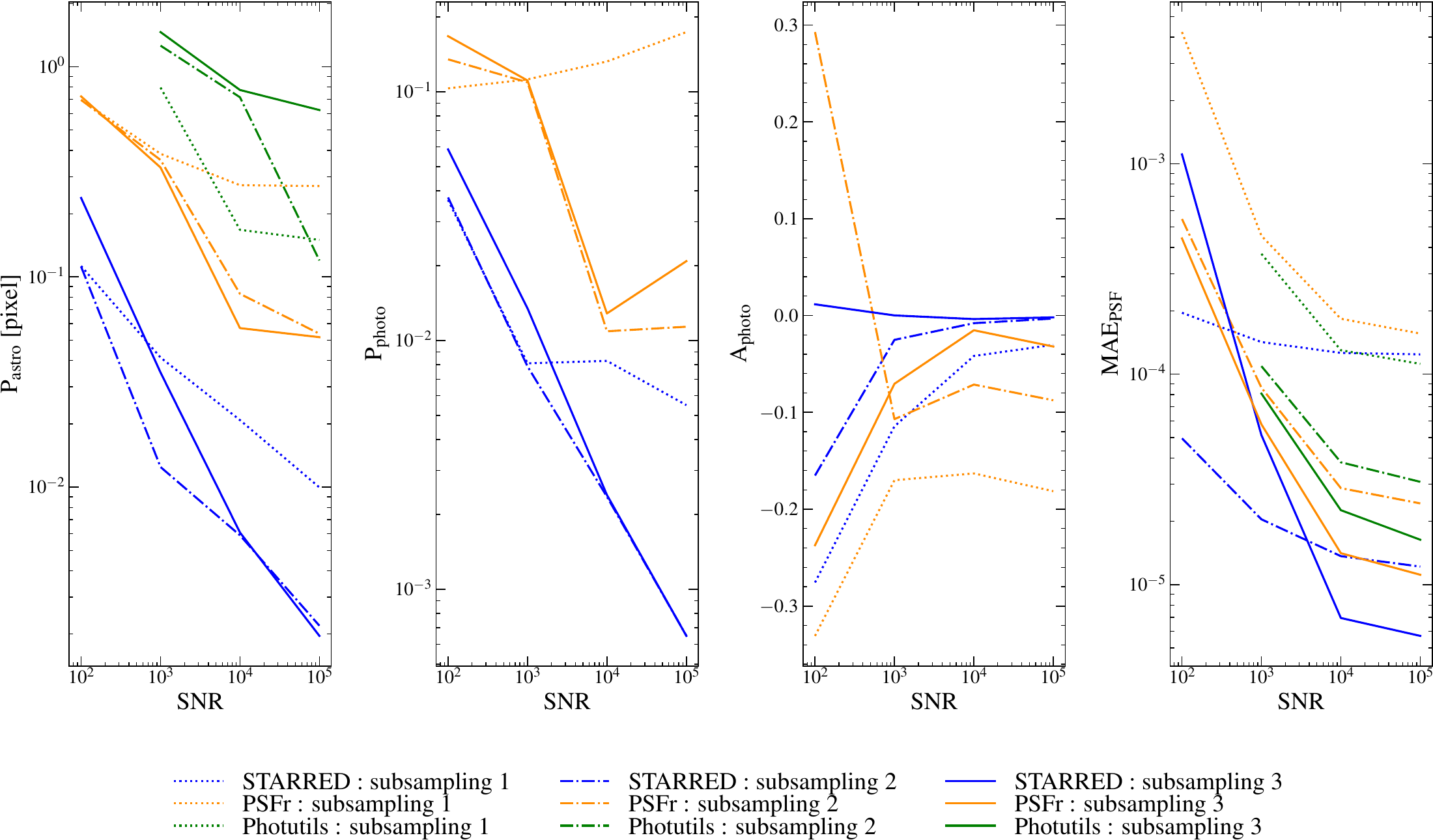}
    \caption{Astrometric precision (first panel), photometric precision (second panel), photometric accuracy (third panel) and PSF reconstruction fidelity (fourth panel) as a function of the SNR of the stars used to reconstruct the PSF. Dotted, dash-dotted and solid lines respectively indicate that the PSF is reconstructed at the same, at twice or three times the original pixel size. \starred's performance (in blue) is compared to that of other empirical PSF reconstruction codes such as \psfr (in orange) and \photutils (in green). The \photutils algorithm failed to converge at a SNR below 100 and the reconstructed PSF models were not sufficiently accurate to extract meaningful PSF photometry, which is why these results are not included in this Figure.}
    \label{fig:metric_plot}
\end{figure*}

We ran 3 different PSF reconstruction algorithms, namely \photutils~\citep{Bradley2023}, \psfr \footnote{\href{https://github.com/sibirrer/psfr}{https://github.com/sibirrer/psfr}. This codes utilizes the features of~\cite{lenstronomy}.} and \starred on this simulated data set and compared the performance in terms of fidelity of the reconstructed PSF, $\mathrm{MAE}_{\rm PSF}$, compared to the true original TinyTim PSF at the same resolution. In all three cases, we performed the reconstruction on a 3-times subsampled pixel grid, corresponding to 0.013\arcsec\ per pixel. The results are shown in Figure~\ref{fig:HST_PSF}. 

In this high-SNR regime, \starred accurately reconstructs the original PSF, especially in the center, where the natural dithering introduced by small and random sub-pixel shifts allows us to retrieve the highest frequencies. Although those frequencies are beyond the Nyquist frequency and are therefore not sampled for a single star, the random dithering between the images allows us to overcome this limit.

In terms of PSF fidelity, we obtain a $\mathrm{MAE}_{\rm PSF}$ of 3.75$ \times 10^{-6}$, $1.25 \times 10^{-5}$ and $1.70 \times 10^{-5}$ for \starred, \psfr and \photutils respectively. This corresponds to an improvement factor of 
3.3 and 4.5 in comparison to \psfr and \photutils.

We repeat this experiment for a range of SNR to evaluate the performance of these methods in different SNR regimes. We used the 4 metrics defined in Section~\ref{subsec:metrics_def} to quantify the astrometric precision, the photometric precision and accuracy and the fidelity of the PSF reconstruction. The results of those tests are summarized in Figure~\ref{fig:metric_plot}. 

In terms of astrometric precision, \starred outperforms \psfr and \photutils by almost an order of magnitude and in all SNR regimes. Notably, \starred achieves this outcome by utilizing the random dithering among the six stars to reconstruct a subsampled version of the PSF, while the sparsity prior helps interpolating on a finer pixel grid. This leads to tight constraints on the position of the stars. In this idealized scenario, where the 6 stars have an identical PSF and differ only in brightness, an astrometric precision better than a hundredth of a (original) pixel is attained for SNR$\geq$10 000 and subsampling factor $\geq$2. In real observations, we expect the precision to be degraded due to PSF field distortion and spectral dependence of the PSF, as discussed in Section \ref{sec:discussion}.

The third panel of Figure \ref{fig:metric_plot} shows the importance of reconstructing a supersampled version of the PSF to obtain unbiased photometric measurements. For SNR$\geq$1000 and a supersampling factor of 3, the photometric accuracy is below 0.3\%, which corresponds to a few milli-magnitude. When the PSF is reconstructed on the same pixel grid as the original data, the photometry is underestimated by 3 to 11\% (for SNR ranging from 1 000 to 100 000). At high SNR, the PSF can be constrained to a finer grid of pixels, leading to a significant gain in terms of astrometric precision, photometric precision and photometric accuracy. However, at low SNR, most algorithms fail at reconstructing the PSF if the supersampling factor is too high. This results in a degradation of the \MAEPSF, as can be seen on the right panel of Figure \ref{fig:metric_plot}.

\section{Deconvolution performance}
\label{sec:deconvolution_performance}
\subsection{Example 1: scene modeling of supernova light curves}
\label{subsec:SNIa}
\paragraph{Time series simulation: } 
In this first test example, we generate mock LSST observations of a Type Ia supernovae (SNIa) exploding in a spiral galaxy. The host galaxy is taken from the HST-COSMOS survey, which we degrade to ground-based resolution by convolving it with a set of empirical PSFs, directly modeled from real monitoring data taken at the 2.2m MPIA telescope in La Silla, Chile~\citep{Millon2020a}. The seeing ranges from 0.61\arcsec\ to 1.47\arcsec\ with a median at 0.92\arcsec\ (FWHM). The mock observations are generated using the image simulation module of {\tt lenstronomy}~\citep{lenstronomy}, assuming a LSST zero point of 28.16 in the r band and a sky brightness of 21.2 magnitude per square arcsecond. We generate 50 epochs of observations at a cadence of 10 days. Each epoch consists of two 30-second visit images separated by 30 minutes (and assumed to have been taken in the same filter) and we introduce dithering randomly distributed around the center of the image and with a standard deviation of 0.3\arcsec. For each epoch, we also generate 6 star cutouts to reconstruct the PSF. The PSF star's R-band magnitude is normally distributed around 16 with a scatter of 0.3 mag. 

\begin{figure*}[h!]
    \centering
    \includegraphics[width = 0.7 \textwidth]{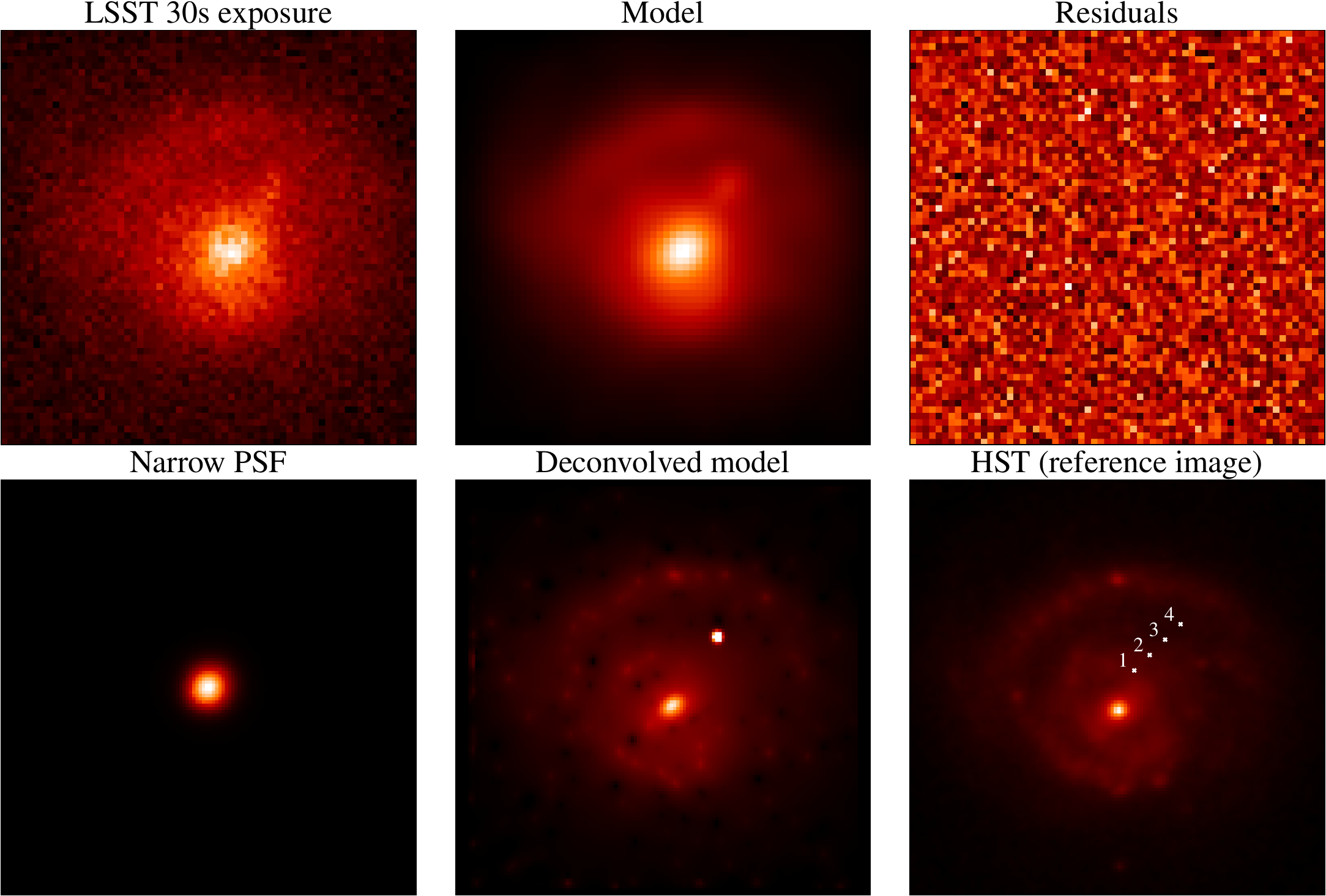}
    \caption{Multi-epoch deconvolution of a galaxy hosting a SNIa. From top to bottom and from left to right, the different panels show a single-epoch image, the model produced by \starred at the original resolution, the residuals for one epoch, the reconstructed narrow PSF \narrowPSFnodep, the deconvolved model and the reference HST image of the host galaxy. In the lower right panel, we indicate the 4 different locations chosen for the SNIa explosion. This example corresponds to location 4, i.e., 2\arcsec\ from the center of the host galaxy. Note that the deconvolved model of the extended channel benefits from the signal-to-noise of the whole dataset.}
    \label{fig:SNIa reconstruction}
\end{figure*}

After the 30$^{\rm th}$ epoch, we inject a varying point source in the galaxy to simulate a SNIa explosion. The light curves are interpolated from {\tt piscola}~\citep{piscola} standard SNIa template and scaled to reach a peak r-band magnitude of 19. We test 4 different locations of the SNIa in the galaxy, marked as white dots on the bottom right panel of Figure \ref{fig:SNIa reconstruction}, to investigate the effect of the local surface brightness on the deblending performance. 

\begin{figure*}[h!]
    \centering
    \includegraphics[width = 0.7 \textwidth]{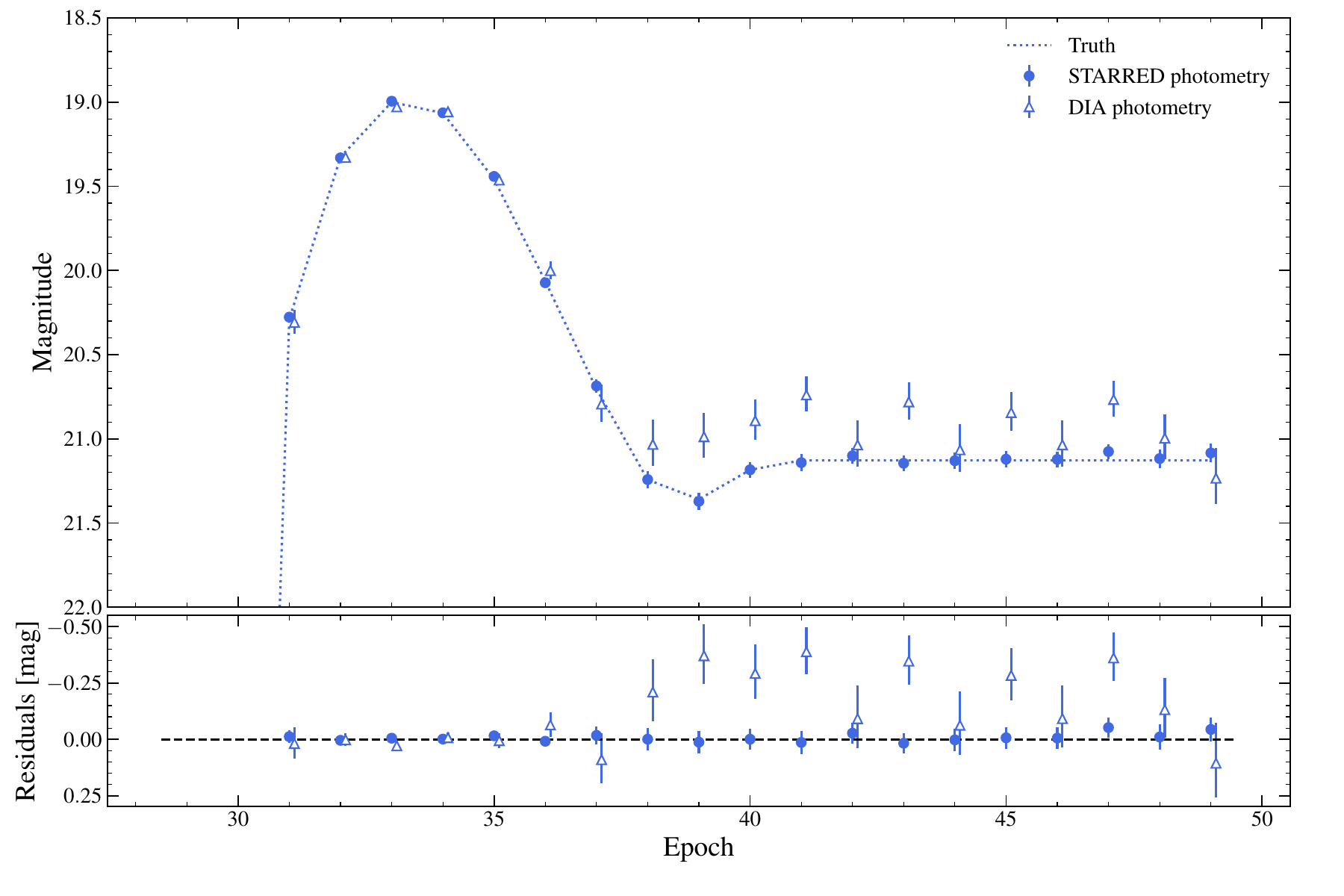}
    \caption{SNIa light curve extracted with \starred using the deconvolved model shown in Figure \ref{fig:SNIa reconstruction} and with DIA + aperture photometry. The bottom panel displays the residuals compared to the true light curve input in the simulation. For the sake of clarity, the 30 observation epochs prior to the SNIa explosion are not shown in this plot. }
    \label{fig:SNIa_lcs}
\end{figure*}

\paragraph{Light curve extraction:} First, we reconstruct the narrow PSF \narrowPSF at each epoch by fitting the 6 star cutouts. Then, we process all 100 images of the SNIa host galaxy simultaneously to reconstruct a high-resolution image of the host galaxy (on a grid of pixel subsampled by a factor of 2 compared to the original data) and to measure the PSF photometry at each epoch. In this example, there is no degeneracy between the point source channel and the extended channel, since we process simultaneously the image before the SNIa explosion. Therefore, $\lambda_{\rm pts}$ is set to 0. We also set $\lambda = 1$, meaning that starlet coefficients 1$\sigma$ above of the noise level are reconstructed in the high-resolution image. This reconstructed high-resolution image is shown in Figure~\ref{fig:SNIa reconstruction} and the extracted PSF photometry is presented in Figure~\ref{fig:SNIa_lcs}. 
For comparison, we also extract the light curve with the difference imaging analysis (DIA) presented in~\cite{Dux2023b}, based on the methodology outlined in~\cite{Alard1998}. 
Difference imaging relies on the subtraction of the flux of all images from a reference epoch.
To achieve perfect subtraction, the resolutions of the images need to coincide, which is achieved by blurring the best seeing image for comparison with every other image. 
With this method, the galaxy is perfectly subtracted, leaving only a single PSF corresponding to the flux excess of the SNIa. The photometry can then be extracted with a simple aperture sum.

\begin{center}
\begin{table*}[h!]
    \begin{tabular}{c|c|ccccc}

			Location & SB [mag.arsec$^{-2}$] & $P_{\rm starred}$ [mag] & $P_{\rm DIA}$ [mag] & $A_{\rm starred}$ [mag] & $A_{\rm DIA}$ [mag] & \Pastro [pixel] \\
			\hline \hline
			1 & 21.931 & 0.029 & 0.145 & -0.037 & -0.111 & 0.083 \\
	
			2 & 22.303 & 0.024 & 0.141 & -0.028 & -0.125 & 0.079 \\
		
			3 & 22.639 & 0.013 & 0.142 & -0.008 & -0.129 & 0.082 \\
	
			4 & 22.808 & 0.013 & 0.144 & 0.001 & -0.129 & 0.081 \\
	
		\end{tabular}
    \caption{Photometric precision, photometric accuracy and astrometric precision recovered by both \starred and difference imaging photometry for 4 different locations of the SNIa in the host galaxy. The light curves are derived from 100 simulated LSST observations of 30 seconds of exposure each. The second column indicates the surface brightness of the host galaxy at the location of the SNIa.}
    \label{tab:SNIa_accuracy}
\end{table*}
\end{center}

\paragraph{Results:}
We use similar metrics to those defined in Section~\ref{sec:PSF_performance} to evaluate the precision and accuracy of the recovered point source photometry of the supernovae. We compute the mean precision and the mean accuracy by comparing, after the supernova explosion, the measured flux to the true flux used in the simulation. We also measure the mean astrometric precision \Pastro by comparing the recovered offset of the image and the true dithering shift input in the simulation. Those metrics are summarized in Table~\ref{tab:SNIa_accuracy}. 

In this example, \starred recovers the position of the supernovae to a precision of about 0.08 pixel, corresponding to 0.017\arcsec. In terms of photometric precision, \starred outperforms the DIA method by a factor of 5 to 10 depending on the SNIa location. The absolute photometry is also less biased, with a mean accuracy ranging from 1 to 37 mmag. The greatest bias occurs when the supernova is closest to the bulge of the lensing galaxy. This level of accuracy is considerably better than that of the DIA method, which typically shows a mean accuracy of 0.11 to 0.13 magnitudes.

\subsection{Example 2: lensed quasar light curves}
\label{subsec:lensed_quasar}
For most lensed quasars, the main factor limiting photometric precision in their light curves is not the photon noise but the deblending errors, most often leading to systematics. Lensed quasars are indeed highly blended objects since their image separation rarely exceeds 2\arcsec, and the light of the lens galaxy and from the gravitational arcs (hereafter \textit{the extended components}) is mixed with the quasar images. Neglecting the blending between the extended components and point sources in lens systems leads to increased scatter in the light curves because these elements are intermixed in a manner influenced by the time-varying (and time-correlated) PSF. The issue is particularly pronounced in systems with small separations between images, where this mixing effect becomes more significant. It can be mitigated, however, by careful modeling of the PSF at each epoch and accurate modeling of the extended components, as we demonstrate below.

Extracting the light curves of lensed quasars is more difficult than in the case of supernovae because the point source does not disappear after a few months. Instead, the point source remains fully degenerate with the background, which makes measuring the magnitude of the quasar extremely challenging. In this example, we test the ability of \starred to lift this degeneracy and obtain precise and accurate light curves of the multiple images of lensed quasars.

\paragraph{Time-series simulation}: We use \lenstro to generate mock images of gravitationally lensed quasars, following a procedure akin to the one described in Section~\ref{subsec:SNIa}, i.e., assuming the same observational and instrumental parameters, including the same empirical PSF as a function of time. We model the lens galaxy as a Sersic profile with Sersic index $n_{Sersic} = 3$ for the light distribution and as a Singular Isothermal Ellipsoid (SIE) for the mass distribution. The source galaxy is modeled as an exponential profile (i.e., $n_{Sersic} = 1$) and a quasar, represented as a point source. We choose a close alignment between the source and the lens to ensure the formation of four multiple images of the quasar, in an Einstein cross configuration. We impose different Einstein radii for the lens, ranging from $\theta_E = 0.5$\arcsec\ to $\theta_E = 1.5$\arcsec\ to test how \starred performs in deblending highly compact lens systems. This range is particularly relevant for galaxy-scale lenses useful for time-delay cosmography, as more compact systems may be difficult to detect.

For the variability of the quasar, we take the real light curve of the lensed quasar RXJ1131$-$1231 as observed by the Euler Swiss telescope in the 2013 season~\citep{Millon2020b}. The light curves are interpolated and rescaled to a mean magnitude of 21.5, before lensing magnification. Since we only intend to test the light curve extraction, we assume that there is no time delay between images, so all images vary simultaneously. This is actually the worst-case scenario as it maximizes possible flux leakage between point sources, making the light curves maximally degenerate. The cadence of observation is set to 10 days.

\begin{figure*}[h]
    \centering
    \includegraphics[width=\textwidth]{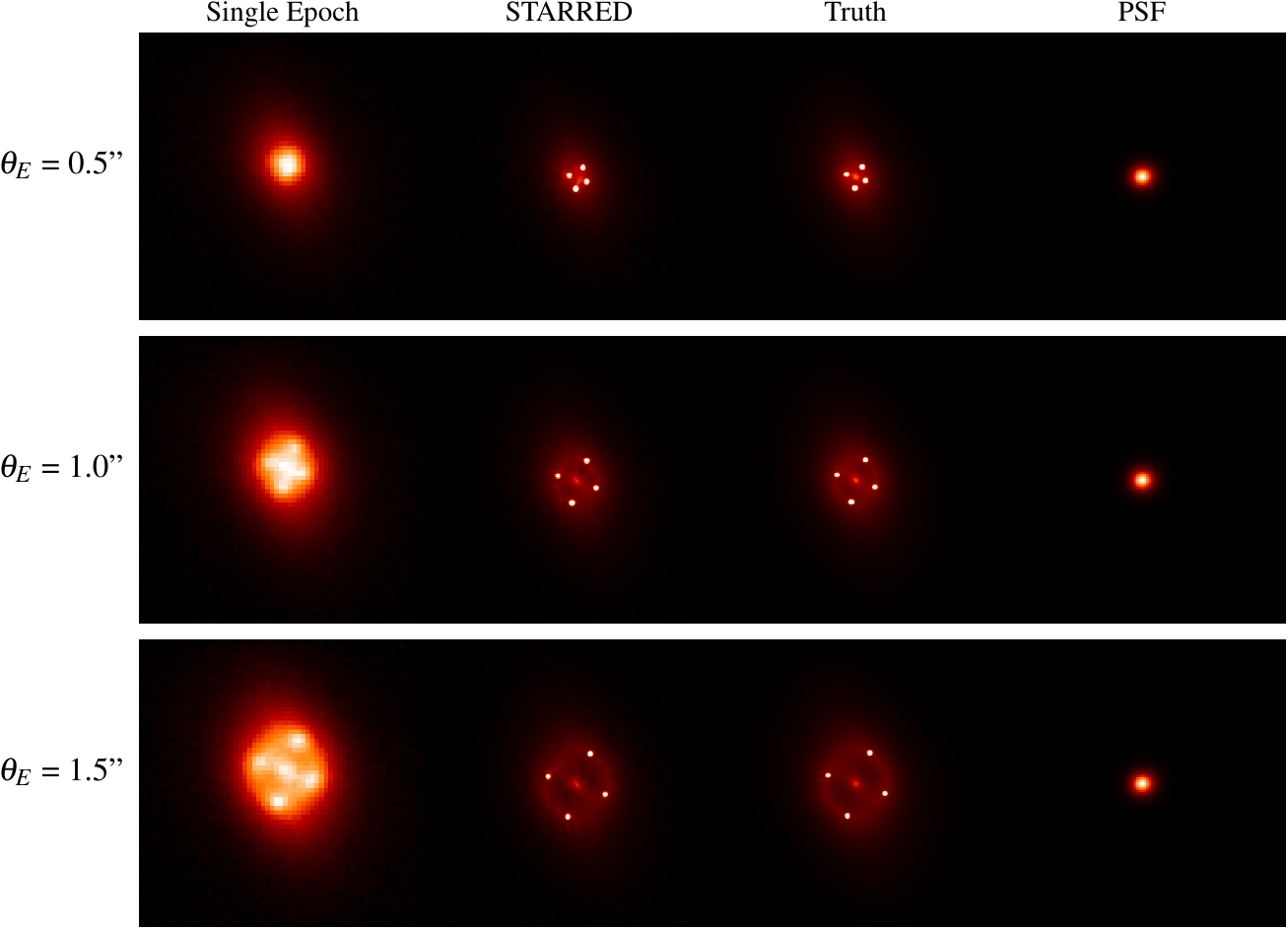}
    \caption{Multi-epoch deconvolution of a quadruply imaged quasar for different Einstein radii $\theta_E$. From left to right, the panels show single-epoch simulated LSST images, the deconvolved models reconstructed from all 100 images, the truth images at the same resolution than the deconvolved image and the PSF.}
    \label{fig:quasar_deconvolution}
\end{figure*}

\begin{figure*}[h]
    \centering
    \includegraphics[width=0.8\textwidth]{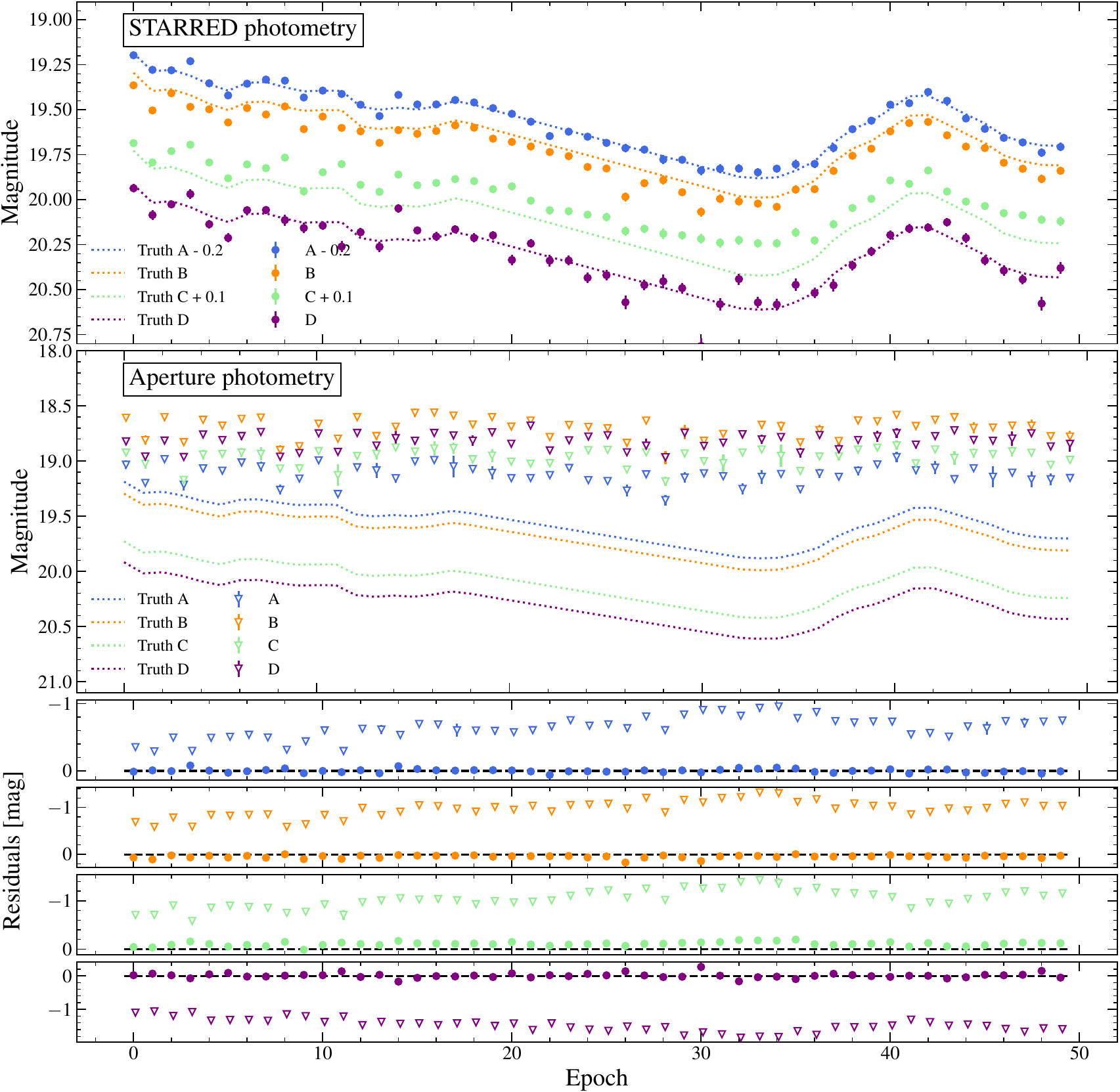}
    \caption{Lensed quasar light curves extracted with \starred using the deconvolved model shown on the second row of Figure \ref{fig:quasar_deconvolution} ($\theta_E$ = 1\arcsec). Aperture photometry obtained after annulus background subtraction is shown with triangles. The bottom panels show the residuals from each lensed image.}
    \label{fig:quasar_lcs}
\end{figure*}

\paragraph{Results:} Using the same reconstructed PSF as in the previous example, we run the deconvolution on all 100 images simultaneously, fitting for both the position and photometry of the multiple images of the quasar. The extended channel is used to reconstruct a model of the light profile of the lens galaxy and of the Einstein ring, at three times the sampling of the original image. The best-fit deconvolved models are shown in Figure~\ref{fig:quasar_deconvolution} and the extracted light curve for each of the four multiple images are shown in Figure~\ref{fig:quasar_lcs}. For comparison, we also extract the photometry with \photutils in an aperture of 8 pixels (1.68\arcsec\ in diameter) for each multiple images. The background is estimated in an annulus of 2 pixels around the aperture. Table~\ref{tab:lQSO_accuracy} provides a summary of the mean photometric precision and accuracy achieved with both \starred and aperture photometry using \photutils.

The recovered mean photometric precision over all 50 observation epochs ranges from 21 to 133 mmag, depending on the image brightness and on the compactness of the lensing configuration. This exhibits a precision approximately two to five times higher compared to the photometry obtained with aperture photometry. In terms of accuracy, the absolute brightness of the images is recovered to less than 0.1 mag in most lensing configurations, except for the most compact one ($\theta_E$ = 0.5\arcsec), where the photometric bias reaches 0.19 mag for image B. This is particularly advantageous, as we intentionally selected a lens system featuring a bright Einstein ring and a bright lens galaxy to assess the deblending capability of \starred. Even if the photometry of one multiple image can be slightly over or under-estimated due to residual deblending errors, the peak in the light curves around epoch 40 is still clearly identified and this does not compromise our ability to measure a time delay. However, for other science cases requiring a measurement of the flux ratio, this systematic error should be carefully quantified and integrated into the overall uncertainty calculations.  

Similar to our previous example, the position of the multiple images is also recovered with a precision of about a tenth of pixel. 

\begin{center}
\begin{table*}[h!]
\resizebox{0.8\textwidth}{!}{
\begin{tabular}{c|c|ccccc}

$\theta_E${[}\arcsec{]} & Image & $\mathrm{P}_{\rm starred}${[}mag{]} & $\mathrm{P}_{\rm aperture}${[}mag{]} & $\mathrm{A}_{\rm starred}${[}mag{]} & $\mathrm{A}_{\rm aperture}${[}mag{]} & \Pastro{[}pixel{]} \\ \hline \hline
\multirow{4}{*}{0.5} & A & 0.076 & 0.149 & 0.089 & -1.526 & 0.134 \\
 & B & 0.076 & 0.162 & 0.187 & -1.771 & 0.174 \\
 & C & 0.077 & 0.150 & -0.002 & -1.837 & 0.097 \\
 & D & 0.133 & 0.166 & 0.009 & -2.543 & 0.364 \\ \hline
\multirow{4}{*}{1.0} & A & 0.021 & 0.129 & -0.002 & -0.638 & 0.089 \\
 & B & 0.024 & 0.136 & 0.050 & -0.966 & 0.092 \\
 & C & 0.031 & 0.149 & -0.108 & -1.037 & 0.085 \\
 & D & 0.049 & 0.150 & 0.007 & -1.472 & 0.078 \\ \hline
\multirow{4}{*}{1.5} & A & 0.023 & 0.149 & 0.028 & -0.622 & 0.086 \\
 & B & 0.024 & 0.121 & 0.102 & -0.605 & 0.088 \\
 & C & 0.026 & 0.150 & -0.061 & -0.965 & 0.076 \\
 & D & 0.037 & 0.154 & -0.049 & -1.182 & 0.075 \\ 
\end{tabular}
}
\caption{Mean photometric precision and accuracy recovered by \starred and aperture photometry (including annulus background subtraction) with \photutils for the simulated LSST images of a quadruply imaged lensed quasar. The last column indicates the mean astrometric precision recovered by \starred.}
\label{tab:lQSO_accuracy}
\end{table*}
\end{center}

\subsection{Example 3: Photometry of high-redshift globular clusters with \textit{JWST}.}
\label{subsec:JWST}

A generic application of \starred is the photometry of point sources superposed on a complex luminous scene. This is well illustrated by the photometry of compact sources associated to the ``Sparkler" galaxy~\citep{Mowla2022} from a single epoch of observation. The ``Sparkler" is a galaxy at redshift z=1.378 strongly lensed by the z=0.39 galaxy cluster SMACS J0723.3-7327. Most of these compact sources (``the sparkles") are unresolved, even with \textit{JWST}, and 5 of them were identified as globular clusters by~\cite{Mowla2022}, owing to their red Spectral Energy Distribution (SED) compatible with the color of old stellar systems. However,~\cite{Mowla2022} used aperture photometry to measure the SED of these sources and most of them appear to be contaminated by the extended galaxy emission. To demonstrate the deblending capability of \starred on real JWST observations, we apply our two-channel deconvolution method, modeling the compact sources in the point source channel and the contaminating light from the galaxy in the extended source channel. We use the reduced NIRCam images of SMACS J0723.3-7327 in F090W, F150W, F200W, F277W, F356W and F444W bands provided by~\cite{Mowla2022}. These images have been reduced with a combination of a modified \textit{JWST} pipeline and the {\tt Grizly} software~\citep{grizly} and have a pixel scale of 0.04\arcsec\ per pixel. All the details about image processing are provided in~\cite{Noirot2023}.

\begin{figure*}[htbp]
    \centering
    \includegraphics[width=0.95\textwidth]{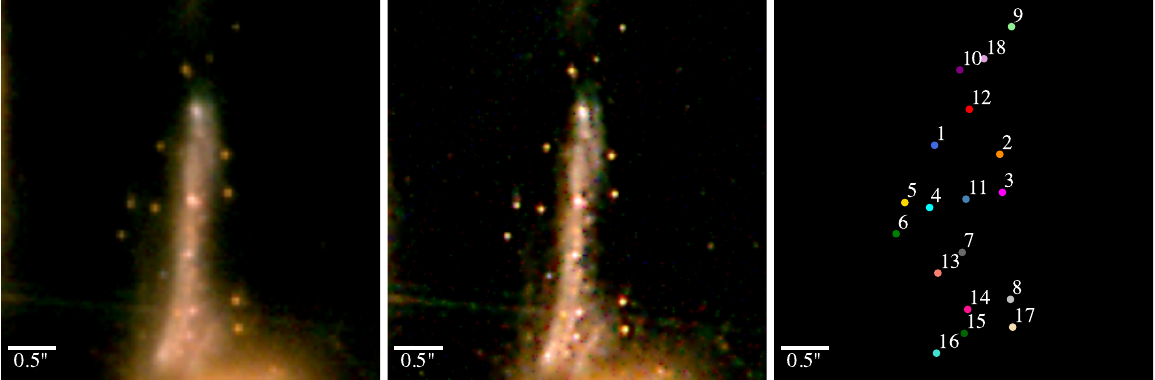}
    \caption{\textit{Left panel:} \textit{JWST}/Nircam color composite (R:F200W, G:F150W, B:F090W) image of the ``Sparkler", a strongly lensed galaxy by the galaxy cluster SMACS J0723.3-7327 (RA=7:23:21.8, DEC=-73:27:18.7). The pixel size is 0.04\arcsec\ per pixel. \textit{Middle panel:} \starred deconvolved model. The pixel size is two time smaller than the original image, i.e., 0.02\arcsec per pixel. \textit{Right panel:} Location and ID number of the ``sparkles", modeled as point sources. }
    \label{fig:sparkler}
\end{figure*}

First, we estimate the narrow PSF in each band using 4 bright stars in the vicinity of the Sparkler. We then use the reconstructed narrow PSFs to perform a deconvolution of the data, improving the sampling by a factor of 2. The reconstructed PSF in the F200W band can be seen on the last panel of Figure~\ref{fig:Wmatrix}. The original and deconvolved images are shown in Figure~\ref{fig:sparkler}, where the identified point sources are labeled from 1 to 18. In addition to the 12 sources measured in~\cite{Mowla2022}, we include 5 additional highly blended sources located in the south of the gravitational arc and a faint one on the north of the arc (\#18). 

\begin{figure*}[htbp]
    \centering
    \includegraphics[width=0.77\textwidth]{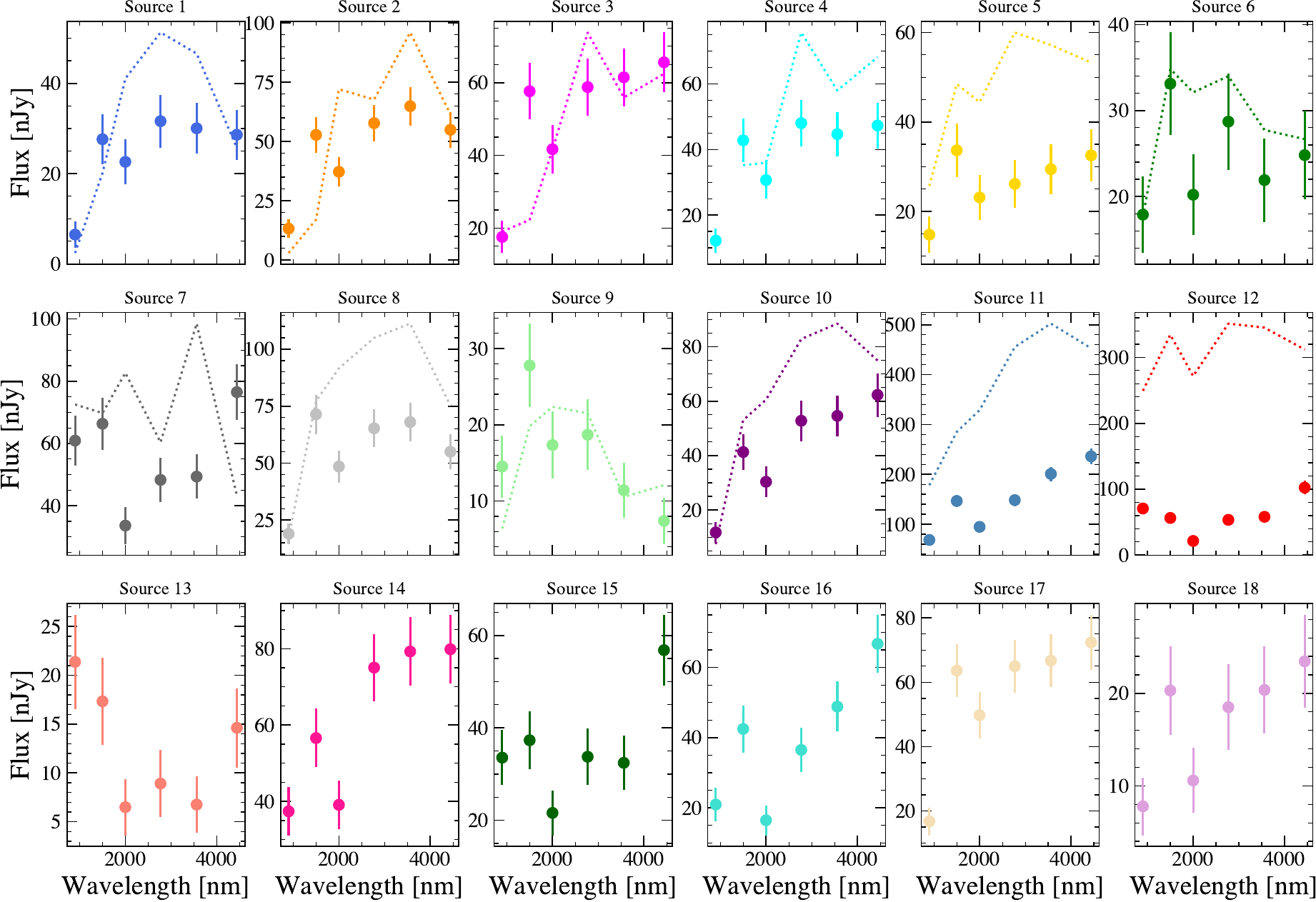}
    \caption{SED of the compact objects labeled in Figure~\ref{fig:sparkler} in the observer frame. The point source photometry is measured from the deconvolved images. Dotted lines correspond to the SED measured in~\cite{Mowla2022} using aperture photometry. Note that point sources \#13 to \#18 were not measured previously. Some of them have SEDs compatible with that of globular clusters such as the ones in \citet{Mowla2022}. In addition, the aperture photometry of point sources exhibiting the closest alignment with the underlying arc is clearly biased compared to \starred.}
    \label{fig:SED_sparkler}
\end{figure*}

\paragraph{Results:} We confirm the point source nature of 13 out of the 18 compact objects modeled in the point source channel. Source ID 5, 6, 7, 11 and 12 are probably extended as some flux residuals are visible in the extended channel near the point sources. From the deblended photometry, we show the SED of these objects in Figure \ref{fig:SED_sparkler}, while Table \ref{tab:SED_sparkler} collects all the measurements. The aperture photometry measured in~\cite{Mowla2022} is shown with dotted lines in Figure~\ref{fig:SED_sparkler}. As our photometric measurements mostly agree for the unblended systems, we observe important differences for the source ID 4, 5, 7, 8, 10, 11, 12. The difference for the most blended system ID 7, 11 and 12 located in the center of the galaxy reach a factor of 2 to 3 difference, highlighting the need of properly modeling the background emission to obtain accurate photometry. Source ID 5 and 6 are extended, which explains the difference between the PSF photometry performed by \starred and the aperture photometry of~\cite{Mowla2022}. For the other point sources, we are able to deblend them from the extended emission for the following two reasons. First, the degeneracy between the extended background and the point sources is not perfect, as the shape of the PSF is known to high precision. Therefore, all the features in the PSF (including the diffraction spikes, the Airy rings, etc...) are used to constrain the amplitude of the point source. Second, the starlet prior is informing the extended source morphology ``under" the point sources. PSF features wrongly attributed to the extended source channel are penalized by this prior.
Some of the newly measured point sources (ID 13 to 18) have similar SEDs than the globular clusters identified in~\citep{Mowla2022} (ID 1, 2, 4, 8 and 10). Future NIRSpec IFU data will confirm (or not) our findings.

\section{Discussion}
\label{sec:discussion}

For time-delay cosmography applications, achieving high photometric precision is essential because it enables the clear identification of features that can be unequivocally matched across the different light curves, facilitating accurate time-delay estimations. While photometric accuracy plays a secondary role for time-delay cosmography, it is critical in microlensing studies. The flux ratios are in fact rarely used for lens modeling but are critical observables when studying microlensing caused by a star in the lens galaxy or millilensing by subhalos intervening along the line of sight~\citep{Schechter2002}. Moreover, imperfect deblending can create misleading contrasts between the time-varying point sources and the static extended sources, leading to false microlensing signals in the differential light curve~\citep[see][for a discussion on this issue]{Sluse2006, Millon2020a}.

Difference imaging and aperture photometry might be a valid competing approach for light curve extraction of the supernovae as these transients eventually disappear, offering the opportunity to measure accurately the surface brightness at their location. However, difference imaging is certainly not a solution for small-separation lensed quasars and supernovae as it requires downgrading the resolution of all but the best-seeing image. In addition, difference imaging still requires to deblend the residuals (i.e., the photometric variations) in each single-epoch residual frame. This can be done for a handful of bright and well-separated lenses \citep[e.g.,][]{Sorgenfrei2024} but the difference imaging method is unsuitable for more compact systems. By contrast, we demonstrated that \starred provides an accurate estimate of point source magnitudes by modeling the extended components of the lens, lifting the remaining degeneracy by applying a morphology prior on the extended sources. This feature can also be useful for single-epoch applications as we have shown in Section~\ref{subsec:JWST}, where we obtained deblended photometry of globular clusters, separating the point-like components from the extended component of the host galaxy. Finally, \starred considered the whole dataset jointly, hence the total signal-to-noise rather than the single-epoch one. \starred can therefore also be seen as way to combine images optimally, correcting for the PSF and naturally drizzling the resulting image.

This deblending capability is especially critical for compact lensed systems ($<$ 1\arcsec\ image separation), which are expected to constitute most of the remaining unknown population since most of the well-separated lensed quasars have already been found in static imaging surveys such as the Dark Energy Survey or Gaia \citep[see e.g., Fig. 19 in][]{Lemon2023}. The temporal dimension provided by the LSST survey leaves considerable hope of detecting these systems through their temporal variability, complementing previous lens searches~\citep[see e.g.,][]{Taak2023}. Turning these systems into useful cosmological probes will require the resolution-improvement capability of \starred to reduce the scatter in the light curves and obtain robust time-delay estimates. 
Moreover, we demonstrated that \starred is capable of recovering the point source positions to a tenth of pixel. This is typically 0.02\arcsec\ for realistic LSST simulated images in average seeing conditions (0.92\arcsec\ FWHM median seeing). This is an important result for many science applications and especially for time-delay cosmography since most of the lensed quasars found in LSST will have limited space-based follow-up to determine the astrometry of the lensed image. Obtaining a precise astrometry from the ground already is crucial for accurate lens modeling and to avoid bias on $H_0$. 

The choice of starlets for the regularization in \starred is well-suited for the representation of generic astronomical objects but it presents limitations for representing Einstein rings. Other families of wavelets, such as curvelets~\citep{Starck2006}, might be a better choice for representing lensed arcs, while \starred is designed to maintain a generic approach to accommodate a wide range of applications. The implementation of alternative regularization schemes for specific applications is left for future work. 

Future improvements also include learning a prior on galaxy morphology using high-resolution imaging surveys like Euclid or Roman. On this matter,~\cite{Adam2023} recently introduced a Bayesian framework that employs score-based likelihood characterization and diffusion models, which allows retrieving posterior samples of deconvolved galaxy light profiles from the data and using a prior learned from real observations. As \starred is implemented in JAX and is auto-differentiable, it would be straightforward to combine it with Machine Learning-based techniques. Promising methods also include hybrid approaches such as Learnlets \citep{Ramzi2023}, which involves learning the dictionary for regularization from actual data. This method aims to merge the strengths of sparsity-based methods, such as exact reconstruction and strong generalization, with the rapid processing and flexible prior learning capabilities of neural networks. 

In its current state, our method does not support multi-band data nor it accounts for PSF field distortion. Addressing the latter would require to implement an iterative correction of the PSF during the deconvolution process or, alternatively, infer the full distortion field during the PSF reconstruction step. 
Finally, incorporating color information could significantly facilitate the separation of point sources from extended background sources, as they often exhibit markedly different colors. This would require to combine \starred with existing sparsity-based deblending tools such as \textsc{muscadet}~\citep{Joseph2016}, which leverages galaxy morphology and color information to perform the deblending of extended sources. 

\section{Conclusion}
\label{sec:conclusion}

In this work, we presented a new PSF reconstruction, deconvolution, and light curve extraction software based on the guiding principles of the MCS/Firedec algorithms. \starred builds on the main concepts of its predecessors, deconvolving images into two distinct channels, one for point sources and the other for extended sources. However, \starred features some major improvements compared to these earlier methods:

\begin{enumerate}
    \item It introduces a new regularization based on starlets, a wavelet family well-suited to represent astronomical objects.
    \item \starred is implemented in JAX and is completely auto-differentiable, ensuring a much faster convergence due to the utilization of gradient-informed optimizers. 
    \item \starred is GPU-compatible, making the code 20 to 30 times faster than its predecessors. It also offers the possibility to process all observation epochs simultaneously, adequately modeling possible shifts and rotations between frames. 
    \item It introduces a new empirical PSF reconstruction technique, also based on sparse regularization. \starred outperforms other empirical PSF reconstruction algorithms such as \photutils or \psfr in terms of the fidelity of the reconstructed PSF, photometric accuracy and astrometric precision. This is due to the sparse regularization, which effectively denoises the reconstructed PSF. \starred leverages the small subpixel shifts between stars relative to the pixel grid to reconstruct a subsampled version of the PSF. Obtaining a subsampled PSF is fundamental given the principle of the method aiming at an improved resolution rather than an infinite one, which requires that the deconvolved model, and therefore the PSF, can be measured on a fine grid of pixel. This is especially true with poorly sampled space data \citep[see e.g.,][]{Symons2021}. We also emphasize that the high-quality PSF reconstruction that STARRED offers has diverse science applications, including lens modelling \citep{Galan2024} or shear measurements \citep{Finner2023}.
\end{enumerate}

As implemented, the method has a wide range of applications as it uses a generic PSF reconstruction algorithm and a robust two-channel deconvolution method. It is optimized for light curve extraction of point-like objects but of course also works with single-epoch observations or without any point sources in image, still providing an optimal way of combining multi-epoch observations with variable PSF.  In this study we limit ourselves to three main science cases using simulated LSST images and real \textit{JWST} observations, arriving to the following findings: 
\begin{itemize}
    \item For transient objects like SNIa, \starred is capable of recovering unbiased photometry by accurately modeling the host galaxy surface brightness. Of note, as the data are deconvolved jointly, the extended channel of the final image has the signal to noise of the whole dataset, hence improving over individual treatment of the single epoch by about square root of the number of epochs.
    \item Light curves of lensed images can be extracted with high photometric precision from LSST data, even for the most compact lensed quasars ($\theta_E <$ 1\arcsec). This might open possible time-delay measurements for an entirely new population of compact lensed systems. 
    \item Point source photometry can be extracted from single-epoch \textit{JWST} observations with \starred, despite the complex PSF of this telescope. Using the Sparkler galaxy as an example, we demonstrated the utility of deblending point-like globular clusters from the extended host galaxy to obtain an unbiased SED.
\end{itemize}

Future enhancements of our method could involve correcting for changes in the PSF across the field of view. This correction would be especially crucial for Adaptive Optics observations, which suffers from significant PSF distortions throughout the field and chromatic dependence \citep{Liaudat2023}. Additionally, deep learning methods may be integrated to learn a more tailored prior on galaxy morphology, improving the generic starlet regularization currently in use. The wavelet basis could be optimized by using Learnlets \citep{Ramzi2023} without losing any generality. By leveraging the abundance of high-resolution data from upcoming space-based surveys like Euclid, future hybrid techniques may provide the best of both worlds: they could learn meaningful priors on galaxy morphology while maintaining the precise reconstruction capabilities inherent to sparsity-based techniques. 

\newpage
\appendix

\section{Sparkler photometry}
\label{ap:A}
\begin{center}
\begin{table}[h]
\resizebox{0.9\textwidth}{!}{
		\begin{tabular}{l|cc|cccccc}
			
			ID & RA & DEC & F090W [nJy] & F150W [nJy] & F200W [nJy] & F277W [nJy] & F356W [nJy] & F444W [nJy] \\
			\hline \hline
			1 & 07:23:21.8648 & -73:27:18.2528 & 6.5 $\pm$ 2.9 & 27.6 $\pm$ 5.5 & 22.6 $\pm$ 4.9 & 31.6 $\pm$ 5.9 & 30.0 $\pm$ 5.6 & 28.6 $\pm$ 5.5 \\
			2 & 07:23:21.6948 & -73:27:18.3440 & 13.3 $\pm$ 3.9 & 52.8 $\pm$ 7.4 & 37.2 $\pm$ 6.2 & 57.7 $\pm$ 7.8 & 64.9 $\pm$ 8.2 & 54.9 $\pm$ 7.5 \\
			3 & 07:23:21.6927 & -73:27:18.7617 & 17.6 $\pm$ 4.4 & 57.6 $\pm$ 7.7 & 41.7 $\pm$ 6.6 & 58.8 $\pm$ 7.8 & 61.5 $\pm$ 7.9 & 65.6 $\pm$ 8.2 \\
			4 & 07:23:21.8770 & -73:27:18.9312 & 12.2 $\pm$ 3.8 & 42.8 $\pm$ 6.7 & 30.7 $\pm$ 5.7 & 48.1 $\pm$ 7.1 & 44.7 $\pm$ 6.8 & 47.3 $\pm$ 7.0 \\
			5 & 07:23:21.9393 & -73:27:18.8505 & 14.8 $\pm$ 4.1 & 33.6 $\pm$ 6.0 & 23.1 $\pm$ 5.0 & 26.2 $\pm$ 5.4 & 29.5 $\pm$ 5.6 & 32.5 $\pm$ 5.9 \\
			6 & 07:23:21.9557 & -73:27:19.1906 & 17.9 $\pm$ 4.5 & 33.1 $\pm$ 6.0 & 20.2 $\pm$ 4.7 & 28.7 $\pm$ 5.6 & 21.9 $\pm$ 4.9 & 24.8 $\pm$ 5.2 \\
			7 & 07:23:21.7966 & -73:27:19.3890 & 60.9 $\pm$ 7.9 & 66.3 $\pm$ 8.3 & 33.6 $\pm$ 6.0 & 48.3 $\pm$ 7.1 & 49.4 $\pm$ 7.1 & 76.5 $\pm$ 8.9 \\
			8 & 07:23:21.6719 & -73:27:19.9046 & 19.0 $\pm$ 4.6 & 71.4 $\pm$ 8.6 & 48.5 $\pm$ 7.1 & 65.3 $\pm$ 8.2 & 68.0 $\pm$ 8.3 & 55.0 $\pm$ 7.5 \\
			9 & 07:23:21.6896 & -73:27:16.9919 & 14.5 $\pm$ 4.1 & 27.8 $\pm$ 5.5 & 17.3 $\pm$ 4.4 & 18.7 $\pm$ 4.6 & 11.4 $\pm$ 3.6 & 7.4 $\pm$ 3.0 \\
			10 & 07:23:21.7977 & -73:27:17.4499 & 11.8 $\pm$ 3.7 & 41.3 $\pm$ 6.6 & 30.4 $\pm$ 5.7 & 52.8 $\pm$ 7.5 & 54.6 $\pm$ 7.5 & 62.3 $\pm$ 8.0 \\
			11 & 07:23:21.7827 & -73:27:18.8342 & 68.4 $\pm$ 8.4 & 146.6 $\pm$ 12.2 & 94.7 $\pm$ 9.8 & 148.4 $\pm$ 12.3 & 200.9 $\pm$ 14.2 & 236.1 $\pm$ 15.4 \\
			12 & 07:23:21.7790 & -73:27:17.8600 & 70.7 $\pm$ 8.5 & 56.4 $\pm$ 7.7 & 21.5 $\pm$ 4.8 & 53.5 $\pm$ 7.5 & 58.1 $\pm$ 7.7 & 102.4 $\pm$ 10.2 \\
			13 & 07:23:21.8462 & -73:27:19.5638 & 21.4 $\pm$ 4.8 & 17.3 $\pm$ 4.4 & 6.5 $\pm$ 2.9 & 8.9 $\pm$ 3.4 & 6.8 $\pm$ 2.9 & 14.6 $\pm$ 4.1 \\
			14 & 07:23:21.7762 & -73:27:20.0096 & 37.4 $\pm$ 6.3 & 56.6 $\pm$ 7.7 & 39.1 $\pm$ 6.4 & 75.1 $\pm$ 8.8 & 79.3 $\pm$ 9.0 & 79.9 $\pm$ 9.0 \\
			15 & 07:23:21.7977 & -73:27:20.2697 & 33.6 $\pm$ 6.0 & 37.3 $\pm$ 6.3 & 21.6 $\pm$ 4.8 & 33.8 $\pm$ 6.0 & 32.4 $\pm$ 5.8 & 56.8 $\pm$ 7.7 \\
			16 & 07:23:21.8571 & -73:27:20.5067 & 20.9 $\pm$ 4.8 & 42.5 $\pm$ 6.7 & 16.5 $\pm$ 4.3 & 36.5 $\pm$ 6.3 & 48.9 $\pm$ 7.1 & 66.7 $\pm$ 8.3 \\
			17 & 07:23:21.6648 & -73:27:20.2113 & 16.8 $\pm$ 4.3 & 63.6 $\pm$ 8.1 & 49.8 $\pm$ 7.2 & 65.0 $\pm$ 8.2 & 66.7 $\pm$ 8.3 & 72.4 $\pm$ 8.6 \\
			18 & 07:23:21.7412 & -73:27:17.3260 & 7.8 $\pm$ 3.1 & 20.3 $\pm$ 4.8 & 10.6 $\pm$ 3.5 & 18.5 $\pm$ 4.6 & 20.4 $\pm$ 4.7 & 23.5 $\pm$ 5.0 \\
		\end{tabular}
         }
	\caption{PSF photometry in the different \textit{JWST}/Nircam bands for 18 compact sources identified around the Sparkler galaxy.}
	\label{tab:SED_sparkler}
\end{table}
\end{center}

\section{Noise propagation in starlet space}
\label{ap:B}

\begin{figure*}[h!]
    \centering
    \includegraphics[width=0.95\textwidth]{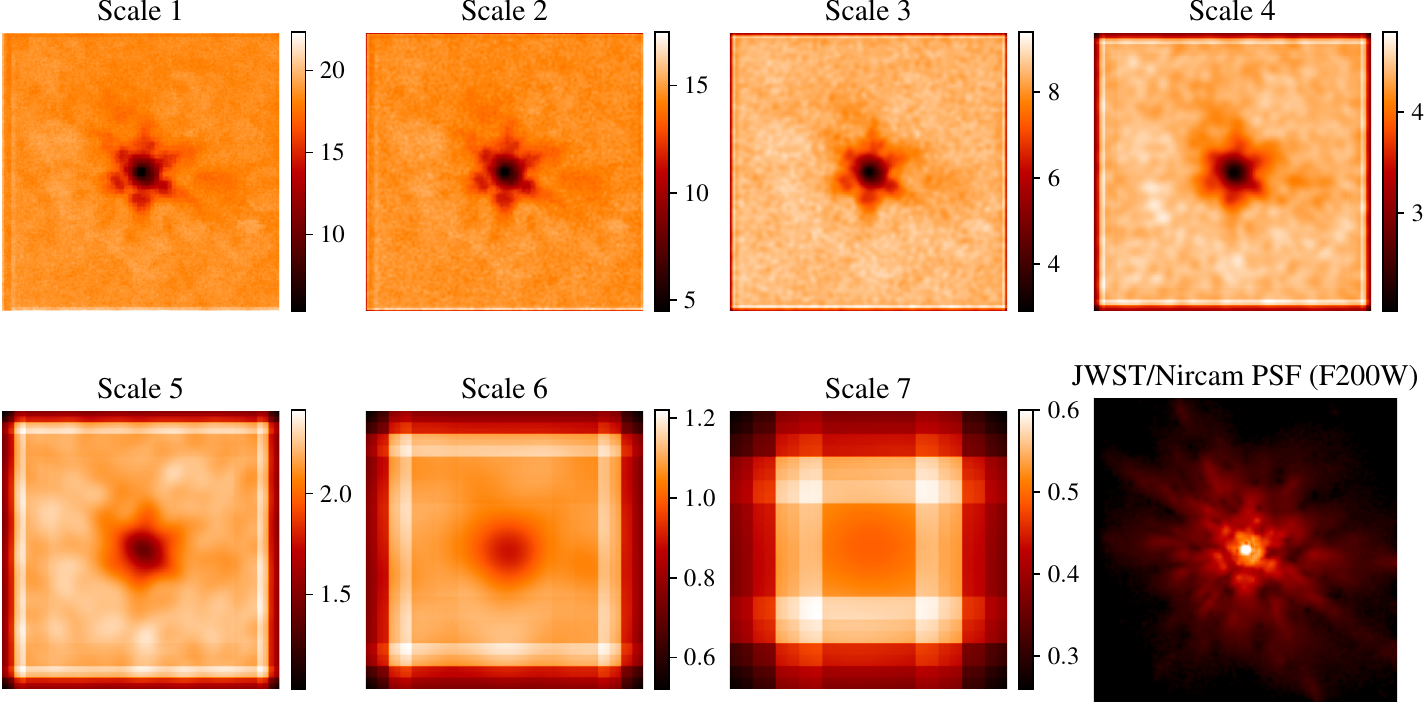}
    \caption{Weight matrix $\weights$ used in the regularization term of Equation~\ref{eq:reg_term}. It is computed in a Monte Carlo procedure by generating 5000 random noise realizations, propagating it into the starlet space and measuring the variance in each starlet scale. In the last panel, we show our reconstructed model of the \textit{JWST}/Nircam PSF in the F200W band at a sampling of 0.013\arcsec\ per pixel. The PSF has been reconstructed from 4 stars located near the ``Sparkler" galaxy in the field of view of the galaxy cluster SMACS J0723.3-732. This Figure can be reproduced from this notebook \href{https://gitlab.com/cosmograil/starred/-/blob/main/notebooks/more_examples/7_JWST-PSF_reconstruction.ipynb?ref_type=heads}{\faGitlab}. 
    }
   \label{fig:Wmatrix}
\end{figure*}

As an illustration of the noise propagation procedure described in Section~\ref{subsec:Noise_prop}, we show in Figure~\ref{fig:Wmatrix} the estimated weight matrix $\weights$ when reconstructing the \textit{JWST}/Nircam F200W PSF used in Section~\ref{subsec:JWST}. 
The first seven panels depict the coefficients of the $\weights$ matrix for each of the seven starlet scales. $\weights$ is used to re-weight the starlet coefficients before computing the $\ell_1$-norm.

\section{Regularization of the point source channel}
\label{ap:C}

The MCS and Firedec algorithms did not incorporate regularization for the point source channel, leading to frequent formation of ``holes" in the extended channel beneath the point sources. This was due to the lack of regularization of the point sources, resulting in them being overweighted in the loss function compared to the extended channel, which was subjected to the sparsity constraint. To address this issue, \starred introduces a correction that compensates for the flux artificially absorbed by the point sources, while keeping a morphology prior on the extended sources.

The correction term $\lambda_{\rm pts}$ in Equation~\ref{eq:argmin_deconv_full} can be estimated from the difference of $\ell_1$-norm of the image with and without the point source channel, normalized by the sum of the $\ell_1$-norm of the individual point sources. The correction factor to apply reads 

\begin{equation}
    \lambda_{\rm pts} = \lambda \frac{\normone{ \weights \odot \waveletop^\top\, \deconvnox} - \lambda\,\normone{ \weights \odot \waveletop^\top\, \hnox}}{ \sum_{k=1}^{M} \langle a_{k} \rangle \normone{ \weights \odot \waveletop^\top\, \bfsym{r}_k}}, 
\end{equation}

where $\deconvnox$ is the average of all single-epoch improved-resolution images $\deconvnox _i$ and  $\langle a_{k} \rangle$ is the mean amplitude of the $k^{\rm th}$ point source. Note that in general
\begin{equation}
\sum_{k=1}^{M} \langle a_{k} \rangle \normone{ \weights \odot \waveletop^\top\, \bfsym{r}_k} \neq \normone{ \weights \odot \waveletop^\top\, \sum_{k=1}^{M} \langle a_{k} \rangle \bfsym{r}_k},
\end{equation}
so the regularization term in Equation~\ref{eq:argmin_deconv_full} does not reduce to $\lambda \normone{ \weights \odot \waveletop^\top\, \deconvnox}$. 

In practice, recomputing the $\ell_1$-norm of $\deconvnox$, $\hnox$ and $\bfsym{r}_k$ at each iteration could lead to convergence problems and be computationally expensive. For that reason, we choose to initially set $\lambda_{\rm pts}$ to $0.5\lambda$ as a first approximation and iteratively refine it from the best-fit model in the previous iteration until convergence is reached.

\begin{acknowledgments}
The authors would like to thank Aymeric Galan for his precious advice on the development of \starred and the CANUCS team for providing us access to their data. MM acknowledges support by the SNSF (Swiss National Science Foundation) through mobility grant P500PT\_203114. KM acknowledges support from the President's PhD Scholarship at Imperial College London. FC acknowledges support from SNSF.
\end{acknowledgments}

\vspace{5mm}

\software{\starred~\citep{Michalewicz2023}, \lenstro~\citep{lenstronomy}, \photutils~\citep{Bradley2023}, \psfr~\citep{lenstronomy}, \jax~\citep{jax}, \textsc{jaxopt}~\citep{jaxopt}, \textsc{optax} \citep{optax}, \textsc{Astropy}~\citep{astropy:2022}.
}

The JWST data presented in this article were obtained from the Mikulski Archive for Space Telescopes (MAST) at the Space Telescope Science Institute. The specific observations analyzed can be accessed via \dataset[DOI: 10.17909/jgpd-kj32]{https://doi.org/10.17909/jgpd-kj32}

\bibliography{main.bib}

\end{document}